 \documentclass[journal]{IEEEtran}
\ifCLASSINFOpdf

\else

\fi

\usepackage{CJK}
\usepackage{algorithm}
\usepackage{algorithmic}
\usepackage{amsmath}
\usepackage{amssymb}
\usepackage{psfrag}
\usepackage{graphicx}
\usepackage{epstopdf}
\usepackage{multirow}
\usepackage{stfloats}
\usepackage{cite}
\usepackage{color}
\usepackage[normalem]{ulem}
\usepackage{arydshln}
\usepackage{enumerate}
\usepackage{bbm}
\newtheorem{lemma}{\bf{Lemma}}[section]

\newcommand{\bm}[1]{\mbox{\boldmath{$#1$}}}

\begin{document}

\title{Channel Estimation for Reconfigurable Intelligent Surface Aided Multi-User mmWave MIMO Systems}

\author{Jie Chen, \IEEEmembership{Member, IEEE}, Ying-Chang Liang, \IEEEmembership{Fellow, IEEE}, \\Hei Victor Cheng, \IEEEmembership{Member, IEEE},  and Wei Yu, \IEEEmembership{Fellow, IEEE}

\thanks{This paper has been accepted in the IEEE Transactions on Wireless Communications.  (Corresponding author: Ying-Chang Liang.)

J. Chen was with the National Key Laboratory of Science and Technology on Communications  and the Center for Intelligent Networking and Communications (CINC), University of Electronic Science and Technology of China (UESTC), Chengdu 611731, China. He is now with the Department of Electrical and Computer Engineering, Western University, London, ON N6A 5B9, Canada (e-mail: chenjie.ay@gmail.com).

Y.-C. Liang was with the Center for Intelligent Networking and Communications (CINC), University of Electronic Science and Technology of China (UESTC), Chengdu 611731, China. He is now with Institute for Infocomm Research (I$^2$R), Agency for Science, Technology and Research (A*STAR), 1 Fusionopolis Way, \#21-01, Connexis South Tower, Singapore 138632, Republic of Singapore (e-mail: liangyc@ieee.org).

H. Cheng  was with the Electrical and Computer Engineering Department, University of Toronto, Toronto, ON M5S 3G4,
Canada. He is now with the Electrical and Computer Engineering Department, Aarhus Universrity, 8200, Aarhus N, Denmark (e-mail: hvc@ieee.org).

W. Yu is with the Electrical and Computer Engineering Department, University of Toronto, Toronto, ON M5S 3G4,
Canada (e-mail: weiyu@comm.utoronto.ca).}

}
\maketitle

\begin{abstract}
Channel acquisition is one of the main challenges for the deployment of reconfigurable intelligent surface (RIS) aided communication systems. This is because an RIS has a large number of reflective elements, which are passive devices with no active transmitting/receiving abilities. In this paper, we study the channel estimation problem for the RIS aided multi-user millimeter-wave (mmWave) multi-input multi-output (MIMO) system. Specifically, we propose a novel channel estimation protocol for the above system to estimate the cascaded channels, which are the products of the channels from the base station (BS) to the RIS and from the RIS to the users.
Further, since the cascaded channels are typically sparse, this allows us to formulate the channel estimation problem as a sparse recovery problem using compressive sensing (CS) techniques, thereby allowing the channels to be estimated with less training overhead. Moreover, the sparse channel matrices of the cascaded channels of all users have a common block sparsity structure due to the common channel between the BS and the RIS. To take advantage of the common sparsity pattern, we propose a two-step multi-user joint channel estimation procedure. In the first step, we make use of the common column-block sparsity and project the received signals onto the common column subspace.
In the second step, we make use of the row-block sparsity of the projected signals and propose a multi-user joint sparse matrix recovery algorithm that takes into account the common channel between the BS and the RIS.
\end{abstract}
% Note that keywords are not normally used for peerreview papers.
\begin{IEEEkeywords}
%\vspace{-0.5cm}
Reconfigurable intelligent surface, multi-user joint channel estimation, compressive sensing.
\end{IEEEkeywords}

\IEEEpeerreviewmaketitle

%\vspace{-0.8cm}
\section{Introduction}
Reconfigurable intelligent surface (RIS), also known as {{intelligent reflective surface}} (IRS), is a promising technique for achieving high spectral and energy efficiencies in wireless systems \cite{yang2016programmable,basar2019wireless,di2019smart,liang2019large,liang2020symbiotic}.
An RIS is composed of a uniform array with a large number of reflective elements, each of which can induce a phase shift on the incident signal and reflect it passively.
By adaptively adjusting the phase shifts of the RIS to allow the reflections to add up constructively, the received power of the intended signal can be enhanced \cite{garcia2019reconfigurable,ellingson2019path,tang2019wireless,Long2020Globecom}. This is also called passive beamforming\cite{wu2019intelligent}.
Compared with traditional {{amplify-and-forward}} (AF) relays \cite{ntontin2019reconfigurable,qingqing2019towards}, RIS elements can reflect the incident signal passively and allow the reflection coefficients to be reconfigured in real-time with little energy consumption \cite{qingqing2019towards} and noise corruption. As a result, an RIS equipped with a large number of reflective elements can provide a high array/passive beamforming gain without requiring much additional hardware cost \cite{chen2019intelligent}.

Due to these promising advantages, RIS has been proposed to be included in various wireless communication systems.
In most of the literature, the main idea is to jointly optimize the beamformer at the transceiver and the phase shift matrix induced by the RIS to achieve various objectives \cite{huang2019reconfigurable,nadeem2019largearxiv,guo2019weighted,pan2019multicell,jung2019optimality,zhou2019intelligent,wang2019intelligent, perovic2019channel,yang2019irs}.
Specifically, for a downlink {{multi-user multiple-input multiple-output}} (MU-MIMO) system
\cite{huang2019reconfigurable,nadeem2019largearxiv,guo2019weighted,pan2019multicell,zhou2019intelligent,jung2019optimality}, the energy-efficiency maximization problem was studied in \cite{huang2019reconfigurable} subject to individual {{Quality-of-Service}} (QoS) constraint. In \cite{nadeem2019largearxiv}, the minimum {{signal-to-interference-plus-noise ratio}} (SINR) maximization problem subject to a maximum power constraint was studied by considering both rank-one and full-rank channel matrices between the {{base station}} (BS) and the RIS.
The weighted sum-rate maximization problem was studied in \cite{guo2019weighted} for a single-cell scenario and in \cite{pan2019multicell} for a multi-cell scenario.
Moreover, the downlink achievable rate maximization problem was studied in \cite{yang2019irs} for a wideband {{orthogonal frequency division multiplexing}} (OFDM) system.  Moreover, the channel capacity optimization problems were studied in \cite{zhou2019intelligent} with single RIS and in \cite{jung2019optimality} with multiple RISs, and then extended to a {{millimeter-wave}} (mmWave) environment in \cite{ wang2019intelligent, perovic2019channel}.

%Furthermore, the achievable sum-rate maximization problem was studied in \cite{zhou2019intelligent} for a single-RIS scenario and in \cite{jung2019optimality} for a multi-RISs scenario.

However, all the above studies  \cite{huang2019reconfigurable,nadeem2019largearxiv,guo2019weighted,pan2019multicell,zhou2019intelligent,jung2019optimality,yang2019irs,perovic2019channel,wang2019intelligent} focus on the joint design of the beamformer at the BS and the phase shift matrix at the RIS under the assumption that {{channel state information}} (CSI) is perfectly known, which is not realistic in practice.
Compared with traditional active devices (i.e., AF relays) aided communication systems, channel estimation
in RIS aided systems is more challenging. This is because in conventional systems with active devices, CSI can be estimated by letting the active devices send training sequences. However, the RIS consists of devices that are passive reflective elements, which cannot perform active transmission/reception. This means that channel estimation can only take place at the BS or at the users, which makes the estimation of channels involving RIS a difficult task \cite{liang2019large}. This motivates us to find innovative channel estimation methods to tackle the new challenges.

Recently, there have been quite a few works \mbox{investigating} the channel estimation problem for the RIS aided \mbox{single} user communication systems \cite{mishra2019channel,he2019cascaded,jensen2019optimal,zheng2019intelligent,you2019intelligent,taha2019enabling,huang2019indoor}. Specifically, the {\mbox{binary}} reflection method was proposed in \cite{mishra2019channel,he2019cascaded}, where the RIS turns on each reflective \mbox{element} {{successively}}, while \mbox{keeping} the remaining reflective \mbox{elements} turned off. Then, the BS {\mbox{successively}} estimates the cascaded channels, which are {\mbox{products}} of the channels from the BS to the reflective \mbox{elements} and from the elements to the users.
In \cite{jensen2019optimal}, a minimum \mbox{variance} unbiased channel estimator was proposed which works without turning off the RIS in the training period. In the same paper, the optimal phase shift matrix for channel estimation (in the sense of minimizing the Cramer-Rao bound) was shown to be a discrete Fourier transform matrix.
This method was further extended in \cite{zheng2019intelligent,you2019intelligent}, where the authors assume that the RIS can be divided into multiple sub-surfaces, and each sub-surface consists of some adjacent reflective \mbox{elements} which share a common reflection coefficient.
\mbox{However}, one of the main problems with the training methods in the literature is that their overheads scale quickly with the number of reflective elements (or sub-surfaces), which makes them less useful in practice.
In \cite{taha2019enabling}, some active elements are randomly deployed at the RIS to perform channel estimation. Then, full CSI can be recovered by using the estimated CSI from the active elements. This method can reduce the training overhead, but it also increases the hardware cost and complexity due to the deployment of active elements on the RIS. In \cite{huang2019indoor}, a { {deep neural network}} (DNN) is designed to learn the mapping between the measured information at a user location and the optimal
configuration of the RIS to maximize its received signal power. In \cite{jiang2021learning}, a DNN that learns the mapping from the received pilot signals to the optimal beamformer at the BS and phase shift configuration at the RIS to maximize the sum-rate is proposed.
However, deep learning approach suffers from generalizability issues, so these methods are not applicable to scenarios where system configuration changes frequently.

Motivated by the above, in this paper, we study the channel estimation problem for the RIS aided multi-user mmWave MIMO communication systems. To highlight the main {\mbox{contributions}}, we summarize the paper as follows:
\begin{itemize}
\item
We propose a novel channel estimation protocol and {\mbox{apply}} {{compressive sensing}} (CS) techniques to estimate the cascaded channels of the RIS aided multi-user mmWave MIMO system.
First, we develop  a sparse representation of the cascaded channels. Since the BS and the RIS are usually mounted at height, there are only limited paths from the BS to the RIS, and from the RIS to the users. This indicates that the cascaded channel has only a few {{angle of arrival}} (AoA) and {{angle of departure}} (AoD) array steering vectors. The sparse channels can be represented by a row-column-block sparse matrix, i.e., there is a two-dimensional sparsity in the channel matrix.
This specific sparsity structure is quite different from the conventional mmWave MIMO communication systems, whose channel matrix usually only has row-block sparsity (i.e., one-dimensional sparsity) \cite{tsai2018efficient,rao2014distributed,ding2018dictionary}.

\item
   Based on the fact that the cascaded channels of all users have the same common BS-RIS channel and different RIS-user channels, we note that all cascaded channels can be normalized with respect to each other using a complex scaling factor. Then, we
   exploit this normalization to come up with a representation that takes into account the characteristics of the common channel explicitly, which enables the development of efficient estimation algorithms.  A similar characterization is further adopted for the wideband system \cite{zheng2020intelligent}.

\item %A conventional way to exploit the common row-column-block sparsity of the channel matrix simultaneously is to quantize the AoAs and AoDs with high resolution, and then apply a sparse recovery algorithm on the discrete grid of AoAs/AoDs. This conventional approach lead to the issues of either large quantization errors or high computational complexity.
    To take advantage of the row-column-block sparsity, i.e., the two-dimensional sparsity, we propose a two-step multi-user joint channel estimation procedure. In the first step, we use a subspace approach and exploit the common column-block sparsity to estimate the common subspace spanned by the AoD array steering vectors. Then, we project the received signals onto this common column (AoD) subspace, which transforms the original row-column-block sparse channel matrix to just a row-block sparsity matrix.
    % Since we use the received signals of all users to recover the sparse matrix jointly, a better recovery performance can be achieved.

 %To avoid the drawbacks of the conventional estimators, we apply the common row-column-block sparsity to jointly estimate the cascaded channels. Specifically, if we recover the sparse matrix by considering the common row-column-block sparsity simultaneously,     we need to quantize the AoAs and AoDs with high resolution to reduce the quantization errors caused by the discrete grid of AoAs/AoDs.     This usually leads to intractable computational complexity. To further deal with this issue, we propose a two-step procedure based multi-user joint channel estimator.     In its first step, we exploit the common column-block sparsity to estimate the common subspace spanned by the AoD array steering vectors. Then, we project the received signals into this common AoD subspace, which can reduce the number of zero columns of the original sparse matrix and transform it as a row-block sparsity matrix. Hence, this procedure can reduce computational complexity due to fewer unknown columns in the sparse matrix, lower quantization error due to not quantizing AoD, and higher SNR due to reducing the influence of noise on the null space of common AoD subspace.     In the second step, we exploit the common row-block sparsity to formulate a MMV-based multi-user joint sparse matrix recovery problem. Since we use the received signals of all users to recover the sparse matrix jointly, a better recovery performance can be achieved.

\item In the second step, we solve the multi-user joint sparse matrix recovery problem which is formulated using our developed model that takes into account the common BS-RIS channel and different RIS-user channels.
    {\mbox{Specifically}}, the model consists of the channel scaling gain and the row-block sparse matrix as coupled variables. This sparse estimation problem can be formulated as an \mbox{optimization} problem using a sparsity promoting log-sum function. We propose an approach based on \mbox{alternative} \mbox{optimization} and iterative reweighted algorithms to solve the problem efficiently. We analyze the convergence, the \mbox{complexity}, and the effects of initializations for the proposed \mbox{algorithm}. Moreover, we design the training sequences of reflection coefficients based on minimizing the mutual coherence of the equivalent dictionary.
    Finally, the \mbox{simulation} results validate the effectiveness of the proposed scheme.%, which outperforms the other estimation schemes significantly.

%Since the optimization variables are coupled in the formulated multi-user joint  sparse matrix recovery problem, which is non-convex and hard to solve, we propose an approach based on alternative optimization and  iterative reweighted algorithms to solve it efficiently. Besides, we analyze the convergence, complexity, and initializations for the proposed algorithm. Moreover, we design a training reflection coefficient sequence optimization method based on minimizing the mutual coherence of the equivalent dictionary.
    %Finally, the simulation results validate the effectiveness of the proposed scheme.%, which outperforms the other estimation schemes significantly.
\end{itemize}

The rest of this paper is organized as follows.
Section \ref{sec:System Model} presents the system model and channel estimation protocol.
Section \ref{sec:ConventionalCS} investigates the sparsity of the cascaded channel, and shows the drawbacks of
the conventional techniques.
{\mbox{Section}} \ref{sec:ProposedCS}  studies the two-step multi-user joint channel {\mbox{estimation}} procedure and Section \ref{sec:SolutionCS} provides its detailed solution.  {Section} \ref{sec:Pilotoptimization} studies the training reflection coefficient optimization method.
Finally, Section \ref{sec:Simulation} provides simulation results and Section \ref{sec:Conclusion} concludes the paper.

{\emph {Notation}}:
The scalar, vector, and matrix are lowercase, bold lowercase, and bold uppercase, i.e., $a$, $ {\bm{a}}$, and $ {\bm{A}}$, respectively.
The transpose, conjugate transpose, trace, and rank are denoted as ${\left(  \cdot  \right)}^{ T}$, ${\left(  \cdot  \right)}^{ H}$, ${\rm{Tr}}\left(  \cdot  \right)$,  and ${\rm{rank}}\left(  \cdot  \right)$, respectively.
The Moore-Penrose pseudoinverse of matrix $\bm A$ is denoted as ${{\bm{A}}^\dag }$.
The $i$-th component of vector $\bm a$ and $i$-th row $j$-th column component of matrix $\bm A$ are denoted as $\left[{\bm a}\right]_i$ and ${\bm A}^{i,j}$, respectively. The linear space spanned by the column vectors of matrix $\bm A$ is denoted as  ${\rm span}({\bm A})$.
We use ${\bm A}^{:,i}$ and ${\bm A}^{:,\Omega}$ to denote the $i$-th column vector of matrix $\bm A$ and the sub-matrix consisting of the columns of matrix $\bm A$ with indices in set $\Omega$, respectively.
We use ${\bm I}_M$  and ${{\bm 1}_{M\times N}}$ to denote the {\mbox{$M$-by-$M$}} identity matrix and  $M$-by-$N$ matrix whose elements are equal to 1.
Finally,  the {circularly symmetric complex Gaussian} (CSCG) distribution with mean $\mu$ and variance $ {\sigma ^2}$ is denoted as ${\cal C}{\cal N}\left( {\mu,{\sigma ^2}} \right)$.%\in {\mathbb C}^{M\times N}

%
%\begin{figure}[htbp]
%\centering
%\begin{minipage}[t]{0.48\textwidth}
%\centering
%   \includegraphics[width = 1\textwidth]{SystemLISCE.eps}
%    \caption{AN RIS aided multi-user MIMO system consisting of one BS with $M$ antennas, one RIS with $L$ reflective elements, and $K$ single-antenna users.}%\vspace{-0.5cm}
%    \label{Fig1SystemModel}
%\end{minipage}
%\begin{minipage}[t]{0.48\textwidth}
%\centering
%  \includegraphics[width = 1\textwidth]{CEProtocol.eps}
%   \caption{Channel estimation protocol and frame structure.}%\vspace{-0.5cm}
%    \label{Fig2Frame}
%\end{minipage}
%\end{figure}

\section{System Model}\label{sec:System Model}

\subsection{Cascaded Channel Model}

\begin{figure}
    \centering
        \includegraphics[width = 0.5\textwidth]{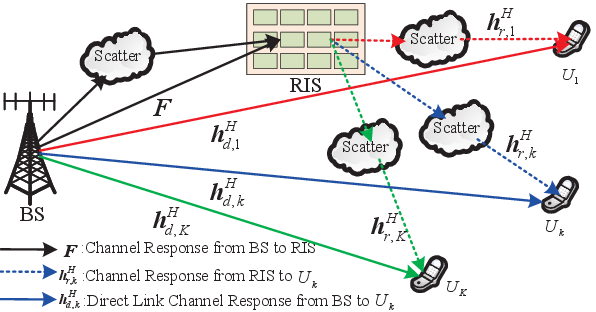}
    \caption{An RIS aided multi-user MIMO system consisting of a BS with $M$ antennas, an RIS with $L$ reflective elements, and $K$ single-antenna users.}
    \label{Fig1SystemModel}
\end{figure}

As shown in Fig.~\ref{Fig1SystemModel}, we consider a block-fading multi-user mmWave MIMO system operating in {{time division duplex}} (TDD) mode, where a BS with the assistance of an RIS serves $K$ users in the same time and frequency. The BS and the RIS are equipped with $M$ antennas and $L$ reflective elements, respectively, while the users are all equipped with a single antenna.  The users are denoted by $U_1,\cdots,U_K$.
%Without loss of generality, we assume $L\ge M \ge K$.
The downlink channel responses from the BS to the RIS, from the RIS to $U_k$, and from the BS to $U_k$ are denoted  by ${\bm F}\in {\mathbb C}^{L \times M}$, ${\bm h}_{r,k}^{H}\in {\mathbb C}^{1 \times L}$, and  ${\bm h}_{d,k}^{H}\in {\mathbb C}^{1 \times M}$, respectively.
The reflective channel at the RIS is usually referred to as the dyadic backscatter channel, where each reflective element combines all the received signals and then reflects them to $U_k$. The reflection matrix \cite{
wu2019intelligent,ntontin2019reconfigurable,qingqing2019towards,chen2019intelligent,huang2019reconfigurable,nadeem2019largearxiv,guo2019weighted} is ${\mathbf V}_{d} ={\text {  diag}}({\bm v})\in {\mathbb C}^{L\times L}$ with ${\bm v} =[v_1,v_2,\cdots,v_L]^{T}\in {\mathbb C}^{L\times 1}$ where $v_l =e^{j\vartheta_l}$ is the coefficient on the $l$-th reflective element.

%\begin{figure}[t]
%\centering
%\begin{minipage}[t]{0.44\textwidth}
%\centering
%   \includegraphics[width = 0.9\textwidth]{SystemLISCE.eps}
%   \caption{ An RIS aided multi-user MIMO system consisting of a BS with $M$ antennas, an RIS with $L$ reflective elements, and $K$ single-antenna users.}
%      \label{Fig1SystemModel}
%\end{minipage}
%  \quad
%\begin{minipage}[t]{0.475\textwidth}
%\centering
%        \includegraphics[width = 0.9\textwidth]{CEProtocol.eps}
%    \caption{{ Channel estimation protocol and frame structure.}}%\vspace{-0.9cm}
%    \label{Fig2Frame}
%\end{minipage}
%\end{figure}

The combined channel response from the BS to $U_k$ through RIS is denoted by ${\bm h}_{d,k}^{H}+{\bm h}_{r,k}^{H}{\mathbf V}_{d}  {\bm F}\in{\mathbb C}^{1\times M}$. Denote the downlink beamformer at the BS for the data signal to $U_k$ by ${\bm w}_k\in{\mathbb C}^{M\times1}$. Then, the SINR of this data signal at $U_k$ scales with
\begin{align}
\!\!\!{\left| {( {{\bm{h}}_{d,k}^H \!+\! {\bm{h}}_{r,k}^H{\mathbf V}_{d}  {\bm{F}}} ){{\bm{w}}_k}} \right|^2}\! \!
= \!\!{\left| {( {{\bm{h}}_{d,k}^H \!+ \!{\bm{v}}^{T}{\rm{diag}}( {{\bm{h}}_{r,k}^H} ){\bm{F}}} ){{\bm{w}}_k}} \right|^2},\label{eq0}
\end{align}
where ${\mathbf V}_{d} ={\text {  diag}}({\bm v})$ and ${\bm{h}}_{r,k}^{H}{\rm{diag}}({\bm{v }}) = {{\bm{v }}^{T}}{\rm{diag}}( {{\bm{h}}_{r,k}^{H}} )$. Please refer to
\cite{chen2019intelligent,wu2019intelligent} for details.

Hence, the transmission quality of the intended incident-reflection signal can be enhanced by adaptively adjusting ${\mathbf V}_{d}$ at the RIS and beamformer ${\bm w}_k$ at the BS, which is called joint active and passive beamforming design. However, such joint design for downlink data transmission requires the CSI of ${\bm h}_{d,k}^{H}$, ${\bm h}_{r,k}^{H}$, and $\bm F$, simultaneously. Moreover, RIS has a large number of reflective elements, which are passive devices without active transmitting and receiving abilities. Hence,   channel estimation in the studied system  is quite a challenging problem.

This paper aims to solve the channel \mbox{estimation} \mbox{problem} for an RIS aided system. Specifically, from \eqref{eq0}, it is \mbox{straightforward} to show that the joint beamformer design only depends on the direct link channel ${\bm h}_{d,k}$  and the following cascaded channel \cite{chen2019intelligent,liang2019large}:
\begin{eqnarray}
{\bm G}_k={\text{diag}}({{\bm{h}}_{r,k}^{H}} ){\bm{F}}\in{\mathbb C}^{L\times M}.\label{eqsignalmodel}
\end{eqnarray}
In the following, we first present an uplink channel estimation protocol, then present the conventional {{least square}} (LS) method to estimate direct channel ${\bm h}_{d,k}$  and cascaded channel ${\bm G}_k$.

\begin{figure}[t]
    \centering
        \includegraphics[width = 0.5\textwidth]{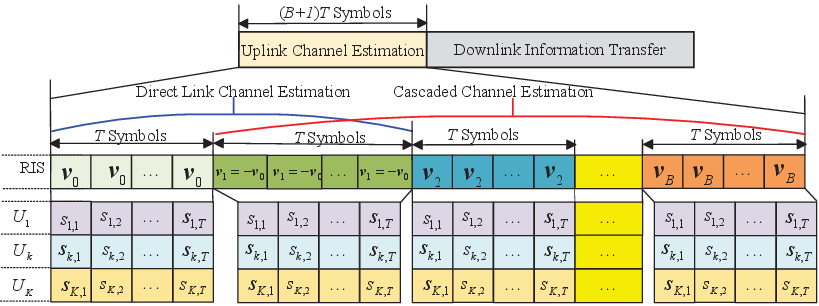}
    \caption{Channel estimation protocol and frame structure.}
    \label{Fig2Frame}
\end{figure}

\subsection{Channel Estimation Protocol}

For TDD systems, the downlink channels are obtained by estimating the uplink channels based on channel reciprocity.
In this paper, we propose the following uplink channel estimation protocol in TDD systems. This protocol is efficient for the case of $L \ge M \ge K$, which holds in most practical scenarios. However, the proposed channel estimator  also works for $M \ge L \ge K$.

To begin with, the channel response from the BS to $U_k$ can be expressed as $ {{\bm{v }}^{T}}{\rm{diag}}( {{\bm{h}}_{r,k}^{H}} ){\bm F}={{\bm{v }}^{T}}{\bm G}_k\in{\mathbb C}^{1\times M}$. In order to recover ${\bm G}_k$ from the received training signals, we need to obtain enough individual observations with different $\bm v$'s so that there are more independent observations than the number of unknowns. Moreover, we let the users transmit orthogonal pilot sequences to simplify the estimation of ${\bm G}_k$ from the received training signals.

Based on the above observations, we propose the pilot and data transmission protocol as shown in Fig. \ref{Fig2Frame}. Specifically, the frame structure is divided into two phases, namely {\mbox{Phase-I}} for uplink channel estimation and Phase-II for downlink data transmission.
In this paper, we only focus on the uplink channel estimation in Phase-I. Channel estimation in Phase-I consists of two sub-frames for direct link channel estimation and $B$ sub-frames for cascaded channel estimation where each sub-frame consists of $T$ time slots with $T\ge K$.
In the proposed protocol, the RIS reflection coefficients ${\bm v}_b=[v_{b,1},v_{b,2},\cdots,v_{b,L}]^{H}\in {\mathbb C}^{L\times 1}$ in different time slots are the same within the same sub-frame $b$, and they are changed to different values in different sub-frames.
The $k$-th user transmits its orthogonal pilot sequence with length $T$ in each sub-frame to the BS, i.e., ${\bm s}_k^{H}=[s_{k,1},s_{k,2},\cdots,s_{k,T}]\in {\mathbb C}^{1\times T}$ with ${\bm s}_{k_1}^{H}{\bm s}_{k_2}=0$ if $k_1\ne k_2$ for $1\le k_1, k_2 \le K$. %Note that the following proposed scheme can be also applied to single-user case.

In the $b$-th sub-frame, the received pilot signals ${{\bm{Y}}_{b}}\in {\mathbb C}^{M\times T}$ at the BS can be written as
\begin{align}
{{\bm{Y}}_{b}} &= \sum\nolimits_{k = 1}^K {\left({\bm h}_{d,k} +{{\bm{F}}^{H}}{\rm diag}({\bm v}_b){{\bm{h}}_{r,k}}\right){{\bm s}_{k}^{H}}}  + {{\bm U}_{b}}\nonumber\\
&{\mathop  = \limits^{({\rm{a}})}} \sum\nolimits_{k = 1}^K {\left({\bm h}_{d,k}+{\bm G}_k^{H}{{\bm{v}}_b}\right){\bm{s}}_k^{H}} + {{\bm U}_{b}}, \label{eqsignal001}
\end{align}
where (a) is due to ${\bm G}_k={\text{diag}}({{\bm{h}}_{r,k}^{H}} ){\bm{F}}$.
Note that ${\bm s}_{k}^{H}{\bm s}_{k}=PT$  is the energy constraint, where $P$ is the transmit power of each user, and ${\bm U}_b\in {\mathbb C}^{M\times T}$ is the  noise matrix, whose elements follow independent identically CSCG with mean zero  and variance ${\delta^2}$.

%In the following parts, we show the details of direct link channel estimation and cascaded channel estimation in Phase-I.

\subsection{Channel Estimator for Direct Link Channel}
In the first two sub-frames of Phase-I, we design the reflection coefficients for the estimation of the direct link channels.
As shown in Fig. \ref{Fig2Frame}, we set ${\bm v}_0=[e^{j\vartheta_{0,1}},e^{j\vartheta_{0,2}},\cdots,e^{j\vartheta_{0,L}}]^{T} $  and ${\bm v}_1=[e^{j({\vartheta_{1,1}+\pi})},e^{j({\vartheta_{1,2}+\pi})},\cdots,e^{j({\vartheta_{1,L}+\pi})}]^{T} $ in sub-frames 0 and 1, respectively. Hence, we have
\begin{align}
&{{\bm{Y}}_{1}}+{{\bm{Y}}_{0}} \nonumber\\
=&2\sum\limits_{k = 1}^K {{\bm h}_{d,k}{\bm{s}}_k^{H}}+ \sum\nolimits_{k = 1}^K {{\bm G}_k^{H}({{\bm{v}}_1}+{{\bm{v}}_0}){\bm{s}}_k^{H}} + {{\bm U}_{0}}+{{\bm U}_{1}}\nonumber \\
=&2{\bm H}{{\bm{ \tilde S}}}+ {{\bm U}_{0}}+{{\bm U}_{1}},
\end{align}
where ${\bm v}_1=-{\bm v}_0$, {${{\bm{ H}}} = [{{\bm{ h}}_{d,1}},{{\bm{h}}_{d,2}},\cdots,{{\bm{h}}_{d,K}}]\in {\mathbb C}^{M\times K}$} and ${{\bm{ \tilde S}}} = [{{\bm{ s}}_{1}},{{\bm{s}}_{2}},\cdots,{{\bm{s}}_{K}}]^H\in {\mathbb C}^{K\times T}$.
Then, we  apply \mbox{traditional} LS channel estimation methods to estimate ${\bm H}$, e.g.,
\begin{align}
{\hat {\bm H}}=\frac{1}{2}{\left( {{{\bm{Y}}_1} + {{\bm{Y}}_0}} \right)}{{\bm{ \tilde S}}}^H\left( {{\bm{ \tilde S}}}{{\bm{ \tilde S}}}^H \right)^{-1}.
\end{align}
Then, we can subtract the contribution of the direct link channel from the received signals in \eqref{eqsignal001} for cascaded channel estimation, i.e.,
\begin{align}
{{{\bm{\mathord{\buildrel{\lower3pt\hbox{$\scriptscriptstyle\frown$}}
\over Y} }}}_b}{\rm{ = }}{{\bm{Y}}_b}{\rm{ - }}{\hat{\bm{H}}}{{\bm{ \tilde S}}} = \sum\nolimits_{k = 1}^K {{\bm G}_k^{H}{{\bm{v}}_b}{\bm{s}}_k^{H}} + {{\bm U}_{b}}+\Delta {{\bm{H}}}{{\bm{ \tilde S}}}, \label{eqsignal1}
 \end{align}
where $\Delta {{\bm{H}}} = {{\bm{H}}} -{\hat{\bm{H}}}$ is the estimation error of ${\bm H}$.

From the above, we see that the direct link channel {\mbox{estimation}} can be performed by properly designing the phase shifts and applying the traditional estimation method. In the following, we assume that the direct link channel can be estimated perfectly, i.e., ${\Delta {{\bm{H}}}}={\bm 0}$. Note that, the estimation error $\Delta {{\bm{H}}}$ adds to the model as extra noise, which only affects the SNR and does not change the procedures and the theory of the proposed estimation methods. Hence, in the rest of the paper, we assume that the direct link channel is known perfectly, and focus on the estimation of the cascaded channel through the RIS.

\subsection{Conventional LS Estimator for Cascaded Channel}
The conventional LS estimator to estimate the cascaded channel is as follows. With the above channel estimation protocol, since ${\bm s}_k^{H}$ in \eqref{eqsignal1} are orthogonal pilot sequences, we have
\begin{eqnarray}
{{\bm{z}}_{b,k}}\buildrel \Delta \over =\frac{1}{{ {PT} }}{{{\bm{\mathord{\buildrel{\lower3pt\hbox{$\scriptscriptstyle\frown$}}
\over Y} }}}_b}{{\bm{s}}_k} = {{\bm{G}}_k^H{{\bm{v}}_b}}  +{\bm u}_{b,k},  \label{eqsinglvector}
\end{eqnarray}
where ${\bm u}_{b,k}=\frac{1}{{ {PT} }} {{\bm{U}}_b}{{\bm{s}}_k} \in {\mathbb C}^{M\times1}$.
Let ${{\bm{  Z}}_k} = [{{\bm{ z}}_{1,k}},{{\bm{ z}}_{2,k}},\cdots,{{\bm{ z}}_{B,k}}]\in {\mathbb C}^{M\times B}$, ${\bm{V}} = [{{\bm{v}}_1},{{\bm{v}}_2},\cdots,{{\bm{v}}_B}]\in {\mathbb C}^{L\times B}$, and ${{\bm{\widetilde U}}_k} = [{{\bm{u}}_{1,k}},{{\bm{u}}_{2,k}},\cdots,{{\bm{u}}_{B,k}}]\in {\mathbb C}^{M\times B}$.
We can rewrite \eqref{eqsinglvector} into the matrix form:
\begin{eqnarray}
{{\bm{ Z}}_k} = {\bm{G}}_k^H{\bm{V}} + {{\bm{\widetilde U}}_k}.\label{eq5}
\end{eqnarray}
Using the conventional LS channel estimator \cite{chen2019Exploiting}, the channel $
{{{\bm{G}}}_k}$ can be estimated by
\begin{eqnarray}
{{{\bm{\hat G}}}_k} ={( {{\bm{V}}{{\bm{V}}}^H})^{ - 1}}{\bm V}{{\bm{ Z}}_k^H} \label{eqLSestimator}.
\end{eqnarray}

It is worth noting that the above LS estimator in \eqref{eqLSestimator} requires that $\bm V$ should be a full rank matrix, which implies $B\ge L$. Thus, it causes large training overhead when the RIS is equipped with a large number of reflective elements. This motivates us to investigate more efficient cascaded channel estimation schemes to reduce the training overhead.%by using the sparsity information hidden in the spatial MIMO channel.

%represented in the virtual angular domain£¬
\section{Cascaded Channel Sparsity Model}\label{sec:ConventionalCS}

In this section, we first develop the sparse representation of the cascade channel for an individual user. Next, we analyze the specific structure of the cascaded channel and the drawbacks of applying conventional CS-based techniques in the studied scenario.
Finally, we study the common sparsity between different users and provide a representation that accounts for the common sparsity explicitly.

\subsection{Individual User Channel Sparsity Representation}
In this section, we investigate the sparse representation of the cascaded channel in the RIS aided communication systems.
Assume the BS and the RIS are each equipped with a {{{\text{uniform} linear array}} (ULA). By applying the physical {\mbox{propagation}} model of wireless channel \cite{tsai2018efficient}, the channels  $\bm F$ and  ${\bm h}_k$ are given by \cite{liu2020matrix}
\begin{align}
{\bm{F}}  \!\!& = \!\sqrt {\frac{{L\!M}}{{N\!_f}}} \!\sum\limits_{p = 1}^{{N_f}}\! {{\alpha _{p}}{\bm{a}}_L\left( \frac{2\varpi}{\rho}{\sin ( \phi _p^{\rm AoA})}\! \right){{\bm{a}}_M^{H}}\left( \frac{2\varpi}{\rho}{\sin ( {\phi _p^{\rm AoD}} )} \!\right)}, \nonumber\\
{{\bm{h}}_{r,k}} \!\! & =\!\sqrt {\frac{{L}}{{N_{h_k}} }}\sum\limits_{q = 1}^{{N_{h_k}}} {{\beta _{k,q}}{\bm{a}}_L\left(\frac{2\varpi}{\rho} {\sin \left( {{\varphi_{k,q}}} \right)} \right)},\label{eqhk}
\end{align}
respectively, where ${\alpha _{p}}$ and ${\beta _{k,q}}$ denote the complex gains of the $p$-th spatial path between the BS and the RIS and the $q$-th spatial path between the RIS and $U_k$, respectively, ${\phi _p^{{\rm{AoD}}}}$ and ${\phi _p^{{\rm{AoA}}}}$ are the {\mbox{$p$-th}} AoD from the BS and  the {\mbox{$p$-t}}h AoA to the RIS, respectively, and ${{\varphi _{q}}}$ is the $q$-th AoD from the RIS to $U_k$. In addition, ${N_f}$ is the number of spatial paths between the BS and the RIS, and $N_{h_k}$ is the number of spatial paths between the RIS and $U_k$, $\varpi $ is the antenna spacing and $\rho $ is the carrier wavelength. We set $\varpi/\rho=1/2$ for simplicity and use $ {{\bm a}_X}\left(\varphi \right)\in {\mathbb C}^{X\times 1}$ to denote the array steering vector with a positive integer $X$, i.e., ${{\bm a}_X}\left( {\varphi } \right)\! = \! \frac{1}{\sqrt{X}}[1,{e^{j\pi \varphi}},\cdots,{e^{j\pi \varphi \left( {X - 1} \right)}}]^{H}.$

From  \eqref{eqhk}, ${\bm G}_k={\text{diag}}({{\bm{h}}_{r,k}^H} ){\bm{F}}$ can be rewritten as
\begin{align}
{{\bm{G}}_k} =& \sum\limits_{p = 1}^{{N_f}} {\sum\limits_{q = 1}^{{N_{{h_k}}}} {( {\sqrt {\frac{{{L^2}M}}{{{N_f}{N_{{h_k}}}}}} {\alpha _p}{\beta _{k,q}}}{{\bm{a}}_L}(\sin (\phi _p^{{\rm{AoA}}}) - \sin ({\varphi _{k,q}})) } }\nonumber \\
  & \left. {{\bm{a}}_M^H(\sin (\phi _p^{{\rm{AoD}}}))} \right).\label{eqsparsitychannel}
\end{align}
Note that ${{\bm{a}}_L}\!( {\sin ( {\phi _p^{{\rm{AoA}}}}  ) \!- \!\sin ( {\varphi _{k,q}} )} )$ is the ($p,q$)-th cascaded AoA/AoD at the RIS/user side, which is called cascaded AoA in the rest of the paper.

To design the CS-based channel estimator, we approximate the cascaded channel in \eqref{eqsparsitychannel}  using the {{virtual angular domain}}  (VAD) representation, i.e.,
\begin{align}
{\bm G}_k={{\bm{A}}_R}{{\bm{X}}_k}{\bm{A}}_T^{H},\label{eqVAD}
\end{align}
where ${{\bm{A}}_R}\in {\mathbb C}^{L\times G_r}$ and ${{\bm{A}}_T}\in {\mathbb C}^{M\times G_t}$ are the dictionary matrices for the angular domain with angular resolutions $G_r$ and $G_t$, respectively, i.e., each column of ${{\bm{A}}_R}$  and ${{\bm{A}}_T}$ represents the array steering vector corresponding to one specific cascaded AoA at the RIS/user side and one specific AoD at the BS side, respectively\footnote{Although ${\sin ( {\phi _p^{{\rm{AoA}}}}  ) \!- \!\sin ( {\varphi _{k,q}} )}\in[-2,2]$, it is straightforward to show that we only need to quantize ${\sin ( {\phi _p^{{\rm{AoA}}}}  ) \!- \!\sin ( {\varphi _{k,q}} )}$ in the domain $[-1,1]$ due to the periodicity of the function $e^{-j\pi({\sin ( {\phi _p^{{\rm{AoA}}}}  ) \!- \!\sin ( {\varphi _{k,q}} )})}$.}, i.e.,
\begin{align}
{{\bm{A}}_R} = \left[ {{{\bm{a}}_L}( { - {1}} ),{{\bm{a}}_L}( { - {1}+ \frac{2}{{{G_r}}}}), \cdots {{\bm{a}}_L}( {{1} - \frac{2}{{{G_r}}}})} \right],\\
{{\bm{A}}_{{T}}} = \left[ {{{\bm{a}}_M}( { - {1}} ),{{\bm{a}}_M}( { - {1} + \frac{2}{{{G_t}}}}), \cdots {{\bm{a}}_M}( {{1} - \frac{2}{{{G_t}}}})} \right].\label{ARdefine}
\end{align}
In addition, ${{\bm{X}}_k}\in {\mathbb C}^{G_r\times G_t}$ is the angular domain sparse matrix, where the non-zero ($i,j$)-th component {\mbox{represents}} the complex gain of the spatial path corresponding to the $i$-th cascaded AoA steering vector at the RIS/user side and the $j$-th AoD steering vector at the BS side.

%\begin{remark}
{\it Remark}: If the RIS is equipped with a $L=L_h\times L_v$ uniform planar array (UPA), we have
\begin{align}
{{\bm{G}}_k} \!&= \!\sum\limits_{p = 1}^{{N_f}} \sum\limits_{q = 1}^{{N_{{h_k}}}} \left({\sqrt {\frac{{{L^2}M}}{{{N_f}{N_{{h_k}}}}}} {\alpha _p}{\beta _{k,q}}}( {{{\bm{a}}_{{L_v}}}( -{{{\tilde \phi }_{p,k,q}}} ) \otimes {{\bm{a}}_{{L_h}}}( {{{\tilde \varphi }_{p,k,q}}} )} )\right.  \nonumber\\
& \qquad\qquad\times \left.{\bm{a}}_M^H\left(\sin (\phi _p^{{\rm{AoD}}})\right)\right),
\end{align}
where ${{\tilde \phi }_{p,k,q}} =  \cos (\phi _p^{\rm{a}})\cos (\phi _p^{\rm{e}}) - \cos (\phi _{k,q}^{\rm{a}})\cos (\phi _{k,q}^{\rm{e}})$ and ${{\tilde \varphi }_{p,k,q}} = \cos (\phi _p^{\rm{a}})\sin (\phi _p^{\rm{e}}) -\cos (\phi _{k,q}^{\rm{a}})\sin (\phi _{k,q}^{\rm{e}})$.
Here, $\phi _p^{\rm{a}}$, $\phi _{k,q}^{\rm{a}}$, $\phi _p^{\rm{e}}$, and  $\phi _{k,q}^{\rm{e}}$ are the corresponding azimuth and elevation angles.
Then, ${\bm G}_k$ can also be written as \eqref{eqVAD} by changing the AoA vectors in the dictionary, i.e., letting ${{\bm{A}}_R} = {{\bm{A}}_{{R_v}}} \otimes {{\bm{A}}_{{R_h}}}$, where ${{\bm{A}}_{{R_v}}} $ and ${{\bm{A}}_{{R_h}}} $ are similar to the definitions in \eqref{ARdefine}. Hence, the following proposed algorithms can also be applied to UPA model directly.
%\end{remark}

\subsection{Sparsity Structure Analysis and Conventional CS-based Techniques}%Then, we analyze the sparsity structure difference between the studied channel and  the conventional mmWave channel.
In this part, we analyze the sparsity structure of the cascaded channel and show the drawbacks of applying the conventional CS-based techniques to recover the sparse matrix ${{\bm{X}}_k}$.

\subsubsection{Sparsity Structure Analysis}
In the studied model, both BS and RIS are mounted at height.  In high frequency bands, e.g., mmWave bands, signals are easily blocked due to the high penetration loss and reduced diffraction effects. Therefore, the numbers of possible paths from the BS to the RIS, and from the RIS to the users, are usually small \cite{liu2020matrix}.
This indicates that there are only a few AoDs and cascaded AoAs, i.e., both $N_f$ and $N_{h_k}$ are small. Hence, ${{\bm{X}}_k}$ has a row-column-block sparsity structure, i.e., there are only a few column/row vectors in ${{\bm{X}}_k}$ being non-zero, as shown in Fig.~\ref{figuretwostep}.
This specific structure of RIS aided communication systems is quite different from the conventional (mmWave) MIMO communication systems, whose sparse channel matrix usually has only row-block sparsity.
This is because there are only limited scatterers at the BS but rich scatterers at the users in the conventional systems \cite{tsai2018efficient,rao2014distributed,ding2018dictionary}.

\subsubsection{Conventional CS-based Techniques}
Substituting the obtained sparsity channel representation of the cascaded channels  \eqref{eqVAD} into \eqref{eq5},  we have
\begin{align}
{{\bm{ Z}}_k^{H}}= {\bm{V}}^{H}{\bm{G}}_k   + {{\bm{\widetilde U}}_k^H} =  {\bm{V}}^{H}{{\bm{A}}_R}{{\bm{X}}_k}{\bm{A}}_T^{H}  + {{\bm{\widetilde U}}_k^H}.\label{eq14}
\end{align}
Hence, we can formulate the channel estimation problem as a sparse channel matrix ${\bm X}_k$ recovery problem using CS-based techniques.
 \begin{figure*}[t]
\centering
\includegraphics[scale=0.6]{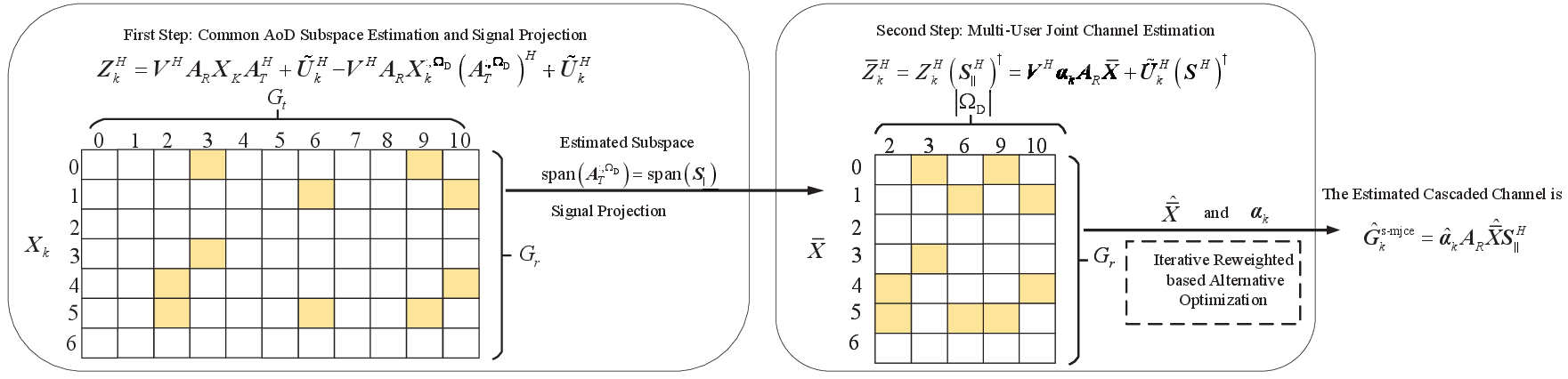}
\caption{An illustration of the sparse pattern in $\mathbf{X}_k$ and the proposed two-step (Subspace) multi-user joint channel estimation procedure (S-MJCE).}
\label{figuretwostep}
\end{figure*}

One straightforward approach is to ignore the block sparsity structure of ${\bm X}_k$, and vectorize  ${{\bm{Z}}_k^{H}}$ and ${\bm X}_k$. The resulting problem can be formulated as a conventional SMV-based problem over an observation vector of size $B \times M$ \cite{tropp2007signal},  i.e.,
\begin{align}
\mathop {\min }\limits_{{{\bm{X}}_k}} & \;{\left\| {{\rm vec}\left( {{{\bm{X}}_k}} \right)} \right\|_0} \nonumber\\
{\rm{s}}{\rm{.t}}{\rm{.}}&\;\left\| {{\rm vec}({\bm{ Z}}_k^H) - \left( {{\bm{A}}_T^{\rm{*}} \otimes {{\bm{V}}^H}{{\bm{A}}_R} } \right){\rm vec}\left( {{{\bm{X}}_k}} \right)} \right\|_2^2 \le \epsilon,
\end{align}
where ${\epsilon}$ is the tolerance upper bound related to the noise, i.e., $\epsilon\ge{\textstyle \frac{MB\delta^2}{PT}}$. This problem can be solved using many different methods, e.g., the {{orthogonal matching pursuit}} (OMP) \cite{rao2014distributed}.

However, the above SMV-based channel estimator has the following disadvantages:
\begin{enumerate}
  \item This estimator requires a larger training overhead to guarantee the estimation performance due to ignoring the block sparsity structure.
  \item It requires super-resolution virtual angle grids for AoAs/AoDs due to the VAD representation in \eqref{eqVAD}. In order to reduce the quantization errors of AoAs/AoDs, both $G_r$ and $G_t$ should be much larger than the number of antennas at the BS or reflective elements at the RIS, thereby resulting in high computational complexity \cite{rao2014distributed}.
\end{enumerate}

Instead of ignoring block sparsity, a better method is to only ignore the column-block sparsity, and use ${{\bm{\widetilde X}}_k}={{\bm{ X}}_k}{\bm{A}}_T^{H}\in{\mathbb C}^{G_r\times M}$ to represent a new sparse channel matrix. Thus, ${{\bm{\widetilde X}}_k}$ has a row-block sparsity structure and can be formulated as a conventional MMV-based recovery problem  over $B$ {\mbox{observation}} vectors of size $M$ \cite{tsai2018efficient,rao2014distributed,ding2018dictionary}, i.e.,
\begin{align}
\mathop {\min }\limits_{{{{\bm{\widetilde X}}}_k}}& \;  {\left\| {{\rm{diag}}\left( {{{{\bm{\widetilde X}}}_k}{\bm{\widetilde X}}_k^{H}} \right)} \right\|_0}\nonumber\\
{\rm{s}}{\rm{.t}}{\rm{.}}&\; \left\| {{\bm{ Z}}_k^{H} - {\bm{V}}^{H}{{\bm{A}}_R}{{{\bm{\widetilde X}}}_k}} \right\|_2^2 \le \epsilon,  \label{eqmmv2}
\end{align}
%where ${\epsilon}$ is the tolerance upper bound related to the noise.
which can be solved efficiently by applying simultaneous OMP (SOMP) algorithm \cite{tsai2018efficient,rao2014distributed}, as shown in Algorithm \ref{algorithm2}.
Particularly, this estimator only needs to recover AoA array steering vectors with ${{\bm{\widetilde X}}_k}$, thus reducing both the {\mbox{dimension}} of the optimization variables and the {\mbox{quantization}} error for AoD representation.
Note that the MMV-based {\mbox{channel}} {\mbox{estimator}} is usually adopted in conventional (mmWave) MIMO {\mbox{communication}} systems, as the channel matrices have the row-block sparsity only.

However, applying  conventional MMV based estimators to the above systems usually leads to a performance loss  because it ignores the specific row-column-block sparsity structure of the cascaded channel. Note that the cascaded channels have the common block sparsity structure brought by the common channel between the BS and the RIS. Thus, recovering the sparse channel matrix of each user individually without considering this common sparsity structure is suboptimal.
This motivates us to redesign the channel estimation procedure.

\begin{algorithm}[t]{\footnotesize
\caption{MMV-based Channel Estimator by using SOMP}\label{algorithm2}
\begin{algorithmic}[1]
\REQUIRE  ~${\bm{ Z}}_k^{H}, 1\le k\le K$, and ${\bm D}={\bm V}^H{{\bm A}_R}$.
\ENSURE ~${\hat {\bm G}}_k^{\rm mmv}$, $1\le k\le K$.
\FOR{$k=1$ to $K$}
\STATE Initialize the residual ${{\bm{R}}_k^{(0)}}={\bm{ Z}}_k^{H}$, the corresponding index set $\Omega_{k}^{(0)}=\emptyset$, and the iteration counter $t=0$.
\REPEAT
\STATE  $t= t + 1$;
 \STATE Estimate
  $ {i_{ k}^{(t)}} = \mathop {\arg \;\max }\nolimits_{i \notin \Omega_{ k}^{(t-1)}} \left\| {( {{\bm{D}}^{:,i}})^{H} {\bm R}_k^{\left( t-1 \right)}} \right\|_2^2$;
 \STATE Update the index set $\Omega_{ k}^{(t)}=\Omega_{ k}^{(t-1)}\cup\{
{i_{ k}^{(t)}}\}$;
 \STATE Update the residual ${{\bm{R}}_k^{(t)}}=( {{\bm{I}} - {\bm{D}}^{:,\Omega_{ k}^{(t)}}{{( {{\bm{D}}^{:,\Omega_{ k}^{(t)}}} )}^\dag }} ){\bm{ Z}}_k^{H}$;
\UNTIL $\left\| {{\bm{R}}_k^{( t )}} \right\|_2^2 \le  \epsilon$.
\STATE ${\hat {\bm G}}_k^{\rm mmv}={{\bm{A}}_R^{:,{\hat \Omega}_{ k}}}{\hat{{\bm{\widetilde X}}}_k}$, where  ${\hat{{\bm{\widetilde X}}}_k}= {{( {{\bm{D}}^{:,{\hat \Omega}_{ k}}} )}^\dag }{\bm{ Z}}_k^{H}$ and ${\hat \Omega}_{ k}=\Omega_{ k}^{(t)}$.
 %Denote the estimated index set by ${\hat \Omega}_{ k}=\Omega_{ k}^{(t)}$, we have ${\hat {\bm G}}_k^{\rm mmv}={{\bm{A}}_R^{:,{\hat \Omega}_{ k}}}{\hat{{\bm{\widetilde X}}}_k}$, where  ${\hat{{\bm{\widetilde X}}}_k}= {{( {{\bm{D}}^{:,{\hat \Omega}_{ k}}} )}^\dag }{\bm{ Z}}_k^{H}$.
\ENDFOR
\end{algorithmic}}
\end{algorithm}

\subsection{Multi-User Common Sparsity Representation}\label{jointsparsity}
Based on the fact that the cascaded channels of all users have the same common BS-RIS channel and different RIS-user channels, we develop a model to represent the common block sparsity of all cascaded channels explicitly. %which together with the row-column-block sparsity will be considered in the design of  multi-user joint channel estimator in Section \ref{sec:ProposedCS}.

\subsubsection{{Common Column-Block Sparsity due to Common {\mbox{Scatterers}} between the BS and the RIS}}

%\item {\bf Joint Column Sparsity due to Common Scatterers at the BS}:
The cascaded
{\mbox{ channels } of all users contain a common component, because the channels from the BS to the RIS are the same. Hence, the AoD array steering vectors of each cascaded channel should be the same as each other.  Specifically, let the AoD index set corresponding to the non-zero columns of ${\bm X}_k$ be ${\Omega _{{{\rm D}},k}}$, where each element is associated with a typical AoD array steering vector at the BS side. Then, from \eqref{eqsignalmodel} and \eqref{eqsparsitychannel}, we have
\begin{align}
 {\Omega _{{{D,}}1}}{\rm{ = }}{\Omega _{{{\rm D,2}}}}\cdots={\Omega _{{\rm D},K}}  \buildrel \Delta \over = {\Omega _{{\rm D}}},\label{eqaodindex}
\end{align}
where $\Omega_{\rm D}$ is defined as the common AoD support index set.

\subsubsection{{Row-Block Sparsity with Joint Scaling Property}}
Since the cascaded AoAs for each user are different, the row-block sparse patterns are different for each user.
Hence, in order to jointly consider the row-block sparsity with the common BS-RIS channel, we need to rewrite the VAD representation of ${\bm G}_k$.
Specifically, we face the following difficulty. Denoting the VAD representation of ${\bm F}$ by ${\bm F}={\bm A}_R{\bm{X}}_F{\bm A}_T^H$,  we  have ${\bm G}_k={\text{diag}}({{\bm{h}}_{r,k}^{H}} ) {\bm A}_R{\bm{X}}_F{\bm A}_T^H$. Then, if ${\bm{h}}_{r,k}^{H}$ is known, we can jointly estimate the common sparsity channel matrix ${\bm{X}}_F$ which is present in all cascaded channels.
However, it is challenging to obtain ${\bm{h}}_{r,k}^{H}$ from ${\bm G}_k$ since there is a scaling ambiguity due to the multiplicative nature of the cascade channel.% between ${\text{diag}}({{\bm{h}}_{r,k}^{H}} )$ and ${{\bm{F}}}$.

To deal with this ambiguity and to develop channel representation, we propose the following scaling property.
Specifically, we observe that the cascaded channels through one arbitrary reflective element for all users are the product of the common channel and different users' channel gains, i.e.,  ${\bm G}_k^{l,:}={{[ {{\bm{h}}_{{{r,k}}}^H} ]}_l}{\bm{F}}^{l,:}$. Hence, each element in the $l$-th row  vector of ${\bm G}_{k_1}$ divided by the corresponding element in the $l$-th row vector of ${\bm G}_{k_2}$ is a common scaling factor, which is equal to  $\frac{{{{[ {{\bm{h}}_{{{r,k}_{\rm{1}}}}^H} ]}_l}}}{{{{[ {{\bm{h}}_{{{r,k}_2}}^H}]}_l}}}$, i.e.,
\begin{equation}
\begin{aligned}
\frac{{{{[ {{\bm{h}}_{{{r,k}_{\rm{1}}}}^H} ]}_l}}}{{{{[ {{\bm{h}}_{{{r,k}_2}}^H}]}_l}}}{\rm{ = }}\frac{{{\bm{G}}_{{k_{\rm{1}}}}^{l,1}}}{{{\bm{G}}_{{k_2}}^{l,1}}}  {\rm{ = }}\frac{{{\bm{G}}_{{k_{\rm{1}}}}^{l,2}}}{{{\bm{G}}_{{k_2}}^{l,2}}} \cdots {\rm{ = }}\frac{{{\bm{G}}_{{k_{\rm{1}}}}^{l,M}}}{{{\bm{G}}_{{k_2}}^{l,M}}}\label{eqscaling}.
% \forall {{k_1}}\forall {{k_2}}, \forall {{l}}\label{eqscaling}.
\end{aligned}
\end{equation}
We call this the joint scaling property, which implies that all cascaded channels can be represented by one arbitrary cascaded channel, and all other channels are scaled versions of it. For example, we can use the cascaded channel of $U_1$, i.e., ${\bm G}_1$, to represent the cascaded channels of the remaining users, i.e.,
  \begin{align}
  {{\bm{G}}_k}& = \underbrace {{\rm{diag}}\left( {{\bm{h}}_{r,k}^H} \right){\rm{diag}}{{\left( {{\bm{h}}_{r,1}^H} \right)}^{ - 1}}}_{{{\bm{\alpha }}_k}}\underbrace {{\rm{diag}}\left( {{\bm{h}}_{r,1}^H} \right){\bm{F}}}_{\bm{G}}
 = {{\bm{\alpha }}_k}{\bm{G}},\label{eqscaling1}
\end{align}
where the subscript ``1'' in ${\bm G}_1$ is omitted for notation simplicity, i.e., ${\bm G}_1={\bm G}$, and
%and ${{\bm{\alpha }}_k} = {\rm{daig}}\left( {{{\left[ {{{\bm{\alpha }}_k}} \right]}_1}, {{\left[ {{{\bm{\alpha }}_k}} \right]}_2},  \cdots {{\left[ {{{\bm{\alpha }}_k}} \right]}_L}} \right) \in {{\mathbb C}^{L \times L}}$, and
%  \begin{equation}
\begin{align}
{{\bm{\alpha }}_k} = {\rm{diag}}\left( \left[\frac{{{{[ {{\bm{h}}_{{{r,k}}}^H} ]}_1}}}{{{{[ {{\bm{h}}_{r,1}^H}]}_1}}}, \frac{{{{[ {{\bm{h}}_{{{r,k}}}^H} ]}_2}}}{{{{[ {{\bm{h}}_{r,1}^H}]}_2}}},  \cdots, \frac{{{{[ {{\bm{h}}_{{{r,k}}}^H} ]}_L}}}{{{{[ {{\bm{h}}_{r,1}^H}]}_L}}} \right]\right) \in {{\mathbb C}^{L \times L}},\label{eqscaling2}% \forall k, \forall l
\end{align}
is referred to as the scaling matrix. We observe from \eqref{eqscaling1} that all cascaded channels have the common channel part ${\bm G}$, which means that all cascaded channels have the following row-block sparse VAD representation, i.e.,
\begin{equation}
\begin{aligned}
{\bm G}_k={{\bm{\alpha }}_k}{{\bm{A}}_R}{{\bm{X}}}{\bm{A}}_T^{H},\label{eqVAD22}
\end{aligned}
\end{equation}
where the subscript ``1'' in ${\bm X}_1$ is omitted for notation {\mbox{simplicity}}, i.e., ${\bm X}={\bm X}_1$.

From \eqref{eqVAD22}, we see that all ${\bm G}_k$ have the common sparse matrix $\bm X$, which is now expressed explicitly.
Note that we use ${\bm{\alpha}}_k{\bm G}$ to represent the cascade channel, because we can use \eqref{eqscaling}  and \eqref{eqscaling1} to separate ${\bm \alpha}_k$ and ${{\bm{G}}}$ from ${\bm G}_k$, thereby resolving the ambiguity in separating ${\text{diag}}({{\bm{h}}_{r,k}^{H}} )$ and ${{\bm{F}}}$ from ${\bm G}_k$.
%This fact is important for the initialization of the multi-user joint channel estimation algorithm studied in the next section.

\section{ Two-Step Multi-User Joint Channel Estimation Procedure}\label{sec:ProposedCS}
\subsection{ Overview of the Two-Step Procedure}

To begin with, it is worth noting that one way to solve the sparse channel recovery problem is to quantize both the AoAs and AoDs, but this would result in a problem with large dimensionality whose the computational complexity would be too high to implement in practice. Observe that the column-block sparsity associated with the AoD subspace is common for all users as the AoD between the RIS and the BS is fixed. In the following, we exploit this property and propose a  two-step (subspace) multi-user joint channel estimation (S-MJCE) procedure, which considers the column sparsity and row sparsity sequentially.

{\bf {First Step}}: By considering the common column-block sparsity, we estimate the common column (AoD) subspace spanned by the AoD array steering vectors, then project the received signals onto this subspace.
Specifically, we have the following operations.
\begin{itemize}
\item We first show that it is sufficient to represent the span of $\bm G_k$ by using a matrix of $N_f$ AoD steering vectors. This   implies that we only need to estimate the common column subspace instead of the exact AoD steering vectors.
\item Then, we solve a maximum likelihood estimation {\mbox{problem}} for subspace estimation.
\item Finally, we project the received signal onto the estimated common column subspace. % to reduce the number of zero columns of ${\bm X}_k$ from $G_t$ to $N_f$.
\end{itemize}

{\bf {Second Step}}: By considering the row-block sparsity with the joint scaling property developed in Section III-C, we formulate a multi-user joint sparse matrix recovery problem.

\subsection{First Step: Subspace Estimation and Signal Projection}\label{subsectionfirst}

To begin with, we denote the AoD matrix, which includes $N_f$ AoD array steering vectors, by ${\bm{A}}_T^{:,{\Omega _{\rm{D}}}}\in {\mathbb C}^{M\times N_f}$, where ${\Omega _{\rm{D}}}$ is the corresponding AoD index set defined in \eqref{eqaodindex}. Then, the common AoD subspace can be represented by the linear span of ${\bm{A}}_T^{:,{\Omega _{\rm{D}}}}$, which is denoted by ${\rm span}({\bm{A}}_T^{:,{\Omega _{\rm{D}}}})$.

Note that if we estimate ${\rm span}({\bm{A}}_T^{:,{\Omega _{\rm{D}}}})$ by estimating the exact $N_f$ AoD array steering vectors in ${\bm{A}}_T^{:,{\Omega _{\rm{D}}}}$ over a discrete grid of AoDs, and then project the signals onto this estimated subspace, it will cause both quantization and estimation errors which   degrades the estimation performance in the second step.

In fact, it is not necessary to estimate the exact $N_f$  AoD vectors in ${\bm{A}}_T^{:,{\Omega _{\rm{D}}}}$  to represent the common AoD subspace ${\rm span}({\bm{A}}_T^{:,{\Omega _{\rm{D}}}})$. We only need to estimate a matrix, ${\bm {S}}_{\parallel} \in {\mathbb C}^{M\times N_f}$, whose linear span includes ${\rm span}({\bm{A}}_T^{:,{\Omega _{\rm{D}}}})$, i.e., finding the columns that contain AoD vectors, i.e.,
\begin{align}
{\bm G}_k&= {{\bm{A}}_R}{{\bm{X}}_k}{\bm{A}}_T^{H}={{\bm{A}}_R}{{\bm{X}}_k^{:,\Omega _{\rm{D}}}}({\bm{A}}_T^{:,{\Omega _{\rm{D}}}})^{H}\nonumber\\
&{\mathop   =  \limits^{({\rm{a}})}}{{\bm{A}}_R}{{\bm{X}}_k^{:,\Omega _{\rm{D}}}}{\bm M}^H{\bm S}_{\parallel}^H{\mathop   =  \limits^{({\rm{b}})}}{{\bm{A}}_R}{{\bm{\bar X}}_k}{\bm S}_{\parallel}^H,\label{eqsubspace2}
\end{align}
where in ({\rm a}) we use the fact that there exists a matrix ${\bm M}\in {\mathbb C}^{ N_f\times N_f}$ such that ${\bm{A}}_T^{:,{\Omega _{\rm{D}}}}  = {\bm {{S}}}_{\parallel}{\bm M}$ if ${\rm span}({\bm{A}}_T^{:,{\Omega _{\rm{D}}}}) \subseteq{\rm span}({\bm {S}}_{\parallel})$, and in ({\rm b}) we use ${{\bm{\bar X}}_k}={{\bm{X}}_k^{:,\Omega _{\rm{D}}}}{\bm M}^H\in{\mathbb C}^{G_r\times N_f}$, which is a row-block sparsity matrix with $N_f$ columns.
Specifically, ${\bm {S}}_{\parallel}$ can be estimated by applying eigenvalue decomposition on the signal covariance matrix. This method does not require quantizing the AoDs. Moreover, we do not need to estimate the linear transform matrix $\bm M$ because that can be included in the combined row-block sparsity matrix ${{\bm{\bar X}}_k}$ in the second step.

The main idea can be seen from the following toy example of \eqref{eqsubspace2}:
\begin{align}
\!\!\!\!\!{{\bm G}_k} &= {{\bm{A}}_R}\underbrace {\left[ {\begin{array}{*{20}{c}}
{{x_{k,11}}\;\;{\rm{0}}\;\;{x_{k,13}}}\\
{{x_{k,{\rm{2}}1}}\;\;{\rm{0}}\;\;{x_{k,{\rm{2}}3}}}\\
{{\rm{0}}\;\;\;\;\;{\rm{0}}\;\;\;\;\;{\rm{0}}}
\end{array}} \right]}_{{{\bm{X}}_k}}\underbrace {\left[ {\begin{array}{*{20}{c}}
{\frac{1}{{\sqrt 2 }}\;\frac{1}{{\sqrt 2 }}\;{\rm{0}}}\\
{\;{\rm{0}}\;\;\;\;{\rm{0}}\;\;\;\;{\rm{1}}}\\
{\;1\;\;\;\;{\rm{0}}\;\;\;\;0}
\end{array}} \right]}_{{\bm{A}}_T^H}\nonumber\\
&= {{\bm{A}}_R}\underbrace {\left[ {\begin{array}{*{20}{c}}
{{x_{k,11}}\;{x_{k,13}}}\\
{{x_{k,21}}\;{x_{k,23}}}\\
{0\;\;\;\;\;0}
\end{array}} \right]}_{{\bm{X}}_k^{:,{\Omega _{\rm{D}}}}}\underbrace {\left[ {\begin{array}{*{20}{c}}
{\frac{1}{{\sqrt 2 }}\;\frac{1}{{\sqrt 2 }}\;{\rm{0}}}\\
{\;1\;\;\;\;{\rm{0}}\;\;\;\;0}
\end{array}} \right]}_{{{({\bm{A}}_T^{:,{\Omega _{\rm{D}}}})}^H}}\nonumber\\
 &= {{\bm{A}}_R}\underbrace {\left[ {\begin{array}{*{20}{c}}
{{x_{k,11}}\;{x_{k,13}}}\\
{{x_{k,21}}\;{x_{k,23}}}\\
{0\;\;\;\;\;0}
\end{array}} \right]}_{{\bm{X}}_k^{:,{\Omega _{\rm{D}}}}}\underbrace {\left[ {\begin{array}{*{20}{c}}
{\frac{1}{{\sqrt 2 }}\;\frac{1}{{\sqrt 2 }}}\\
{\;\;0\;\;\;\;{\rm{1}}\;}
\end{array}} \right]}_{{{\bm{M}}^H}}\underbrace {\left[ {\begin{array}{*{20}{c}}
{0\;\;\;\;1\;\;\;\;0}\\
{\;1\;\;\;\;{\rm{0}}\;\;\;\;0}
\end{array}} \right]}_{{\bm{S}}_\parallel ^H}\nonumber\\
&= {{\bm{A}}_R}\underbrace {\left[ {\begin{array}{*{20}{c}}
{\frac{{{x_{k,11}}}}{{\sqrt 2 }}}&{\frac{{{x_{k,11}}}}{{\sqrt 2 }} + {x_{k,13}}}\\
{\frac{{{x_{k,21}}}}{{\sqrt 2 }}}&{\frac{{{x_{k,21}}}}{{\sqrt 2 }} + {x_{k,23}}}\\
0&0
\end{array}} \right]}_{{{{\bm{\bar X}}}_k}}\underbrace {\left[ {\begin{array}{*{20}{c}}
{\begin{array}{*{20}{c}}
0&1&0
\end{array}}\\
{\begin{array}{*{20}{c}}
1&0&0
\end{array}}
\end{array}} \right]}_{{\bm{S}}_\parallel ^H}.\label{linearspan}
\end{align}
From \eqref{linearspan}, it is obvious that  ${\rm span}({\bm{A}}_T^{:,{\Omega _{\rm{D}}}}) \subseteq{\rm span}({\bm {S}}_{\parallel})$.
Hence, we can reduce the number of columns by projecting the signals onto the estimated subspace ${\rm span}({\bm {S}}_{\parallel})$ without estimating the exact AoD array steering vectors in ${\bm{A}}_T^{:,{\Omega _{\rm{D}}}}$.

More generally, we propose the following procedure to estimate the common AoD subspace from the received signals \eqref{eqsignal1}. The procedure is based on finding the largest eigenmodes of the covariance matrix of the received signal.

To begin with, by substituting ${\bm G}_k ={{\bm{A}}_R}{{\bm{X}}_k^{:,\Omega _{\rm{D}}}}({\bm{A}}_T^{:,{\Omega _{\rm{D}}}})^{H} $ into \eqref{eqsignal1} and ignoring the direct link channel estimation error $\Delta {{\bm{H}}}$, we have
\begin{align}
{{{\bm{\mathord{\buildrel{\lower3pt\hbox{$\scriptscriptstyle\frown$}}
\over Y} }}}_b}&= {\bm{A}}_T^{:,{\Omega _{\rm{D}}}}
{\sum\nolimits_{k = 1}^K {{\left( {{\bm{X}}_k^{:,{\Omega _{\rm{D}}}}} \right)^H}{\bm{A}}_R^H{{\bm{v}}_b}{\bm{s}}_k^{H}}} + {{\bm U}_{b}}. \label{eqsignal10}%\nonumber &={\bm{A}}_T^{:,{\Omega _{\rm{D}}}}{{\bm{P}}_{b}}+ {{\bm U}_{b}},
 \end{align}

Denote
\begin{align}
{\bm{Y}}=\left[ {{{{\bm{\mathord{\buildrel{\lower3pt\hbox{$\scriptscriptstyle\frown$}}
\over Y} }}}_1},{{{\bm{\mathord{\buildrel{\lower3pt\hbox{$\scriptscriptstyle\frown$}}
\over Y} }}}_2},\cdots,{{{\bm{\mathord{\buildrel{\lower3pt\hbox{$\scriptscriptstyle\frown$}}
\over Y} }}}_B}} \right]\in{\mathbb C}^{M\times BT} \label{eqsignal10v2}.%\nonumber &={\bm{A}}_T^{:,{\Omega _{\rm{D}}}}{{\bm{P}}_{b}}+ {{\bm U}_{b}},
 \end{align}
%{\textcolor{red}{If we know $N_f$ in advance and  ${{{\bf{P}}_b}}$ follows complex Gaussian distribution, we have the following lemma for estimating the common AoD subspace.}
If we know the value of $N_f$ in advance, we have the following lemma for estimating the common AoD subspace.
\begin{lemma}\label{lemma01}
For a given $N_f$, by maximizing the likelihood function of $\bm Y$ associated with the AoD vectors in ${\bm{A}}_T^{:,{\Omega _{\rm{D}}}}$, the optimal $N_f$ basis vectors of the estimated subspace is
\begin{align}
{\bm S}_{\parallel}&= \left[ {{{\bm{S}}^{:,1}},{{\bm{S}}^{:,2}},\cdots,{{\bm{S}}^{:,N_f}}} \right],\label{eqsubspace00}
\end{align}
where  ${{\bm{S}}{\bm\Theta} {{\bm{S}}^{{H}}}}$ is the eigenvalue decomposition of ${\bm {\hat C}}=\frac{1}{{BT}}{\bm{Y}}{{\bm{Y}}^H}$, and ${{\bm\Theta} }={\rm diag}([{\theta _1},{\theta _2},\cdots,{\theta _{M}}])\in{\mathbb C}^{M\times M}$ is the eigenvalue matrix where the eigenvalues $\theta_m$ are ordered in a decreasing order in magnitude.
 \end{lemma}
 \begin{IEEEproof}
Please refer to the Appendix.
\end{IEEEproof}

%From Appendix \ref{appendixvA}, we know ${\rm span}({\bm{A}}_T^{:,{\Omega _{\rm{D}}}}) \subseteq{\rm span}({\bm {S}}_{\parallel})$, which is the estimated subspace consisting of $N_f$ vectors.
%{\textcolor{red}{
%In practice, although ${\bm P}_b$ may not follow the Gaussian distribution, the proposed
%method can still achieve good performance from the simulation results.}
In practice, $N_f$ is unknown and has to be estimated. The estimation of $N_f$ is a model selection problem \cite{akaike1998information}. More complex models (with larger $N_f$) can lead to smaller fitting error, but can also overfit. Finding the optimal $N_f$ requires a carefully designed regularization to minimize the fitting error while avoiding overfitting. This issue can be addressed by many well-known decision rules, such as %likelihood ratio \cite{paulraj199316},
{{Akaike's information criterion}} (AIC) and {{ minimum description length}} (MDL) \cite{wax1985detection}.
Based on experimental results, MDL criterion \cite{wax1985detection} has the best performance in our problem. With MDL criterion, the estimate of $N_f$ is given by
\begin{align}
 {{\hat N}_f} =& \mathop {\arg \min }\limits_{{n}} \left\{ { - \log {{\left(  {\frac{{\prod\nolimits_{i = n + 1}^M {\theta _i^{\frac{1}{{M - n}}}} }}{{\frac{1}{{M - n}}{\sum\nolimits_{i = n + 1}^M {{\theta _i}} }}}} \right)}^{ ( {M - n} )BT}}}\right. \nonumber\\
&\qquad\qquad\qquad\qquad\left.{+  \frac{1}{2}n( {2M -n} )\log(BT)} \right\}.\label{eqnumberaod1}
\end{align}

Once the subspace is determined by using eigenvalue decomposition, as next step,  we project the received signals onto the common subspace ${\rm span}({\bm {S}}_{\parallel})$. From \eqref{eq14} and \eqref{eqsubspace2}, we have
\begin{align}
{{\bm{\bar Z}}_k^{H}} \buildrel \Delta \over = {{\bm{ Z}}_k^{H}}({\bm S}_{\parallel}^H)^\dag
=&{\bm{V}}^{H}{{\bm{A}}_R}{{\bm{\bar X}}_k}{\bm S}_{\parallel}^H({\bm S}_{\parallel}^H)^\dag   + {{\bm{\widetilde U}}_k}({\bm S}_{\parallel}^H)^\dag  \nonumber\\
{\mathop   =  \limits^{({\rm{a}})}} & {\bm{V}}^{H}{{\bm{A}}_R}{{\bm{\bar X}}_k}  + {{\bm{\widetilde U}}_k^H}({\bm S}_{\parallel}^H)^\dag ,\label{eqVAD04}
\end{align}
%\begin{align}
%{{\bm{\bar Y}}_k^{H}} \buildrel \Delta \over = {{\bm{\widetilde Y}}_k^{H}}({\bm S}_{\parallel}^H)^\dag
%& ={\bm{V}}^{H}{{\bm{A}}_R}{{\bm{\bar X}}_k}{\bm S}_{\parallel}^H({\bm S}_{\parallel}^H)^\dag   + {{\bm{\widetilde U}}_k}({\bm S}_{\parallel}^H)^\dag \nonumber\\
%&{\mathop   =  \limits^{({\rm{a}})}}  {\bm{V}}^{H}{{\bm{A}}_R}{{\bm{\bar X}}_k}  + {{\bm{\widetilde U}}_k}({\bm S}_{\parallel}^H)^\dag ,\label{eqVAD04}
%\end{align}
where ({\rm{a}}) is due to ${\bm S}_{\parallel}^H({\bm S}_{\parallel}^H)^\dag={\bm I}_{N_f}$.

Comparing with \eqref{eq14} and \eqref{eqVAD04}, we see that the noise power is  reduced from ${\textstyle {\mathbb E}( {\| {{\bm{\widetilde U}}_k} \|_F^2} )}={\textstyle {MB\delta^2}/{PT}}$ to  ${\textstyle {\mathbb E}( {\|  {{\bm{\widetilde U}}_k}({\bm S}_{\parallel}^H)^\dag\|_F^2} )= {N_fB\delta^2}/{PT}}$ due to the removal of noise components in the null space of the common AoD subspace. This leads to a higher effective SNR.
Further, since the projection procedure reduces the number of columns of the sparse matrix, it also reduces the complexity of subsequent row-sparse recovery operation.

\subsection{ Second Step: Multi-User Joint Sparse Matrix Recovery}\label{secproblem}

By the joint scaling property \eqref{eqVAD22}, we can define ${{\bm{\bar X}}} $ as the combined common row-block sparse matrix. Then, the projected signals in \eqref{eqVAD04} can be rewritten as
\begin{align}
{{\bm{\bar Z}}_k^{H}}%& = {\bm{V}}^{H}{{\bm{\alpha }}_k}{{\bm{G}}}({\bm S}_{\parallel}^H)^\dag  + {{\bm{\widetilde U}}_k}({\bm S}_{\parallel}^H)^\dag \nonumber\\
&=  {\bm{V}}^{H}{{\bm{\alpha }}_k}{{\bm{A}}_R}{{\bm{\bar X}}}  + {{\bm{\widetilde U}}_k^H}({\bm S}_{\parallel}^H)^\dag.\label{eqVAD5}
\end{align}
%\begin{align}
%{{\bm{\bar Y}}_k^{H}} &= {\bm{V}}^{H}{{\bm{\alpha }}_k}{{\bm{G}}}({\bm S}_{\parallel}^H)^\dag  + {{\bm{\widetilde U}}_k}({\bm S}_{\parallel}^H)^\dag\nonumber \\
%& =  {\bm{V}}^{H}{{\bm{\alpha }}_k}{{\bm{A}}_R}{{\bm{\bar X}}}  + {{\bm{\widetilde U}}_k}({\bm S}_{\parallel}^H)^\dag.\label{eqVAD5}
%\end{align}
In order to estimate ${{\bm{\alpha }}_k}$ and the real cascaded AoAs
{\mbox{associated}} with the sparse channel matrix ${{\bm{\bar X}}}$ from \eqref{eqVAD5},  we can
{\mbox{formulate}} the following MMV-based multi-user joint sparse matrix recovery problem, i.e.,
%\begin{align}
%\!\!\mathop {\min }\limits_{{{{\bm{\bar X}}}},{\bm \alpha}_k}{\!\!\left\| {{\rm{diag}}\left( {{{{\bm{\bar X}}}}{\bm{\bar X}}_0^{H}} \right)} \right\|_0}\!\!=\!\!\sum\limits_{i = 1}^{{G_r}} {{{\left\| {{\bf{\bar x}}_i^H{{{\bf{\bar x}}}_i}} \right\|}_0}}
%\;\;\;{\rm s.t.}  \left\| {{\bm{\tilde Y}}_k^{H} - {\bm{V}}^{H}{{\bm{\alpha }}_k}{{\bm{A}}_R}{{{\bm{\bar X}}}}} \right\|_2^2 \le {\bar \epsilon}, 1\le k\le K, \label{eqmmv3}
%\end{align}\;
\begin{align}
\mathop {\min }\limits_{{{{\bm{\bar X}}}},{\bm \alpha}_k}&{\left\| {{\rm{diag}}\left( {{{{\bm{\bar X}}}}{\bm{\bar X}}^{H}} \right)} \right\|_0}=\sum\limits_{i = 1}^{{G_r}} {{{\left\| {{\bm{\bar x}}_i^H{{{\bm{\bar x}}}_i}} \right\|}_0}}\;\nonumber\\
{\rm s.t.}\;\;  &\left\| {{\bm{\bar Z}}_k^{H} - {\bm{V}}^{H}{{\bm{\alpha }}_k}{{\bm{A}}_R}{{{\bm{\bar X}}}}} \right\|_2^2 \le {\bar \epsilon},\; 1\le k\le K, \label{eqmmv3}
\end{align}
where ${{\bm{\bar x}}_i^H}$ is the $i$-th row vector of ${{{\bm{\bar X}}}}$, and ${\bar\epsilon}\ge{\textstyle  \frac{N_fB\delta^2}{PT}}$ is the tolerance upper bound related to the combined noise. %The value of $\varsigma$ can be related to \cite{fang2016super} for details.

Denoting the solutions of problem \eqref{eqmmv3} as ${\bm{\hat {\bar X}}}$ and ${\bm {\hat \alpha}}_k$, the cascaded channels can be recovered by
\begin{equation}
\begin{aligned}
{\hat {\bm G}}_k^{\rm  s-mjce}={\bm {\hat \alpha}}_k{{\bm{A}}_R}{\bm{\hat {\bar X}}}{\bm{S}}_\parallel^{H}.\label{eqchannel1}
 \end{aligned}
\end{equation}
Algorithm~\ref{algorithmSub}  summarizes the overall  two-step  multi-user joint channel estimation procedure.

\begin{algorithm}[t]{\small
\caption{Two-Step (Subspace) Multi-User Joint Channel Estimation Procedure (S-MJCE)}\label{algorithmSub}
\begin{algorithmic}[1]
\REQUIRE  ~~${\bm Y}, {\bm{ Z}}_k^{H}, 1\le k\le K$, ${\bm A}_R$, and $\bm V$.
\ENSURE  ~~${\hat {\bm G}}_k^{\rm s-mjce}$, $1\le k\le K$.
\STATE {\bf First Step: Subspace Estimation and Signal Projection}\\{
{\bf $\bullet$ Subspace Estimation}\\
  Compute the covariance matrix ${\bm {\hat C}}=\frac{1}{{BT}}{\bm{Y}}{{\bm{Y}}^H}$; \\
  Compute the eigenvalue decomposition of ${\bm{\hat C}}$ as ${\bm{\hat C}}={{\bm{S}}}{\bm\Theta}{\bm{S}}^{H}$; \\
  Estimate the number of AoD's, ${\hat N}_f$, using \eqref{eqnumberaod1};\\
   The AoD subspace is estimated to be the linear span of the first ${\hat N}_f$ columns of $\bm S$ in \eqref{eqsubspace00}.\\
{\bf $\bullet$  Signal Projection}\\
%Project signals as ${{\bm{\widetilde Y}}_k^{H}}({\bm S}_{\parallel}^H)^\dag = {\bm{V}}^{H}{{\bm{A}}_R}{{\bm{\bar X}}_k}  + {{\bm{\widetilde U}}_k}({\bm S}_{\parallel}^H)^\dag$ in \eqref{eqVAD04}}.
Project the received signals onto the estimated common AoD subspace to obtain \eqref{eqVAD5}.}
 \STATE {\bf  Second Step: Multi-User Joint Sparse Matrix Recovery}
 \\ Solve problem \eqref{eqmmv3} using Algorithm \ref{algorithm3} to obtain $\bm {\hat \alpha}_k$ and ${\bm{\hat {\bar X}}}$;\\
  The cascaded channel is estimated by
${\hat {\bm G}}_k^{\rm  s-mjce}={\bm {\hat \alpha}}_k{{\bm{A}}_R}{\bm{\hat {\bar X}}}{\bm{\hat S}}_\parallel^{H}.$
\end{algorithmic}}
\end{algorithm}

It remains to solve the non-convex problem \eqref{eqmmv3}, where the optimization variables ${{{\bm{\bar X}}}}$ and ${\bm \alpha}_k$ are coupled. In the next section, we propose an algorithm to iteratively optimize ${{{\bm{\bar X}}}}$ and ${\bm \alpha}_k$.

\section{Solution to the Multi-User Joint Sparse Matrix Recovery Problem}\label{sec:SolutionCS}

 In this section, we develop a method based on \mbox{alternating} optimization and iterative reweighted algorithms to solve
{\mbox{problem \eqref{eqmmv3}} efficiently.
%\section{Subspace-based Iterative Reweighted Alternative Optimization for  Multi-User Joint Channel Estimation}

\subsection{Algorithm Development}
%To deal with the coupled optimization variables ${{{\bm{\bar X}}}}$ and ${\bm \alpha}_k$ in \eqref{eqmmv3},

%In this section, we develop an alternative optimization method to solve the multi-user joint  sparse matrix recovery problem \eqref{eqmmv3}.

To begin with, notice that the $l_0$-norm  appearing in \eqref{eqmmv3} is a discontinuous function, a common practice is to replace $l_0$-norm with a continuous function, such as $l_1$-norm, this is so called a surrogate. More recently, \cite{candes2008enhancing,IterativeReweightedWipf2010} show both theoretically and experimentally that using a log-sum function as surrogate is superior to using the $l_1$-norm for sparse signal recovery. Hence, we replace the $l_0$-norm with a log-sum function and rewrite \eqref{eqmmv3} as
\begin{align}
\mathop {\min }\limits_{{{{\bm{\bar X}}}},{\bm \alpha}_k} \;&{\cal Q}({{{\bm{\bar X}}}})=\sum\limits_{i = 1}^{{G_r}} {\log \left( {{\bm{\bar x}}_i^H{{{\bm{\bar x}}}_i} + \varsigma } \right)}\nonumber\\
{\rm s.t.}\;\;& \left\| {{\bm{\bar Z}}_k^{H} - {\bm{V}}^{H}{{\bm{\alpha }}_k}{{\bm{A}}_R}{{{\bm{\bar X}}}}} \right\|_2^2 \le {\bar\epsilon},1\le k\le K,  \label{eqmmv4}
\end{align}
where $\varsigma>0$ is a small positive parameter to ensure the argument inside the $\log$ function is strictly positive. The choice of $\varsigma$ is discussed in \cite{fang2016super}, interested readers are referred there for more details.

Next, by introducing a non-negative penalty factor $\lambda_k$, we reformulate problem \eqref{eqmmv4} as the following unconstrained optimization problem,
\begin{align}
\mathop {\min }\limits_{{{{\bm{\bar X}}}},{\bm \alpha}_k}& \;{\cal L}({{{\bm{\bar X}}}},{\bm \alpha}_k)
\buildrel \Delta \over =  {\cal Q}({{{\bm{\bar X}}}}) + \sum\limits_{k = 1}^K \lambda_k\left\| {{\bm{\bar Z}}_k^{H} - {\bm{V}}^{H}{{\bm{\alpha }}_k}{{\bm{A}}_R}{{{\bm{\bar X}}}}} \right\|_2^2. \label{eqmmv5}
\end{align}
Note that $\lambda_k$ is the penalty factor balancing the tradeoff between data fitting and the sparsity of the solutions. The choice of $\lambda_k$ depends on the power level and is set as $\lambda_k=\frac{PTd}{\delta^2 \log{Gr}}$ \cite{fang2016super,carrillo2009iteratively,tipping2001sparse}, where $d$ needs to be determined based on experimental results. Since the received signal for each user has the same power level, we set $\lambda=\lambda_1=\lambda_2=\cdots=\lambda_K$ throughout the paper.
If the users have different path losses, i.e., they have different power levels,
we can normalize ${{\bm{ z}}_{b,k}}$ with $\textstyle{\frac{{{\bm{  z}}_{b,k}}}{{{\mathbb E}\left( {\beta _{_{k,q}}^2} \right)}}}$, where ${\beta _{_{k,q}}}$ defined in \eqref{eqhk} is the complex gain of the $q$-th spatial path between RIS and $U_k$, and the expectation is taken with respect to the small scale fading. Thus, the normalized received signals have the same power level and our method can still be applied.

Next, to deal with the coupled optimization variables ${{{\bm{\bar X}}}}$ and ${\bm \alpha}_k$, we apply alternating optimization to decouple problem \eqref{eqmmv5}  into the following two unconstrained subproblems
\begin{equation}
\begin{aligned}
\mathop {\min }\limits_{{\bm \alpha}_k}& \;{\cal L}({{{\bm{\bar X}}}},{\bm \alpha}_k),\label{eqmmv51}
\end{aligned}
\end{equation}
which is an optimization problem in ${\bm \alpha}_k$ for a given ${{\bm{\bar X}}}$, and
\begin{equation}
\begin{aligned}
\mathop {\min }\limits_{{{\bm{\bar X}}}}& \;{\cal L}({{{\bm{\bar X}}}},{\bm \alpha}_k),\label{eqmmv52}
\end{aligned}
\end{equation}
which is an optimization problem in ${{\bm{\bar X}}}$ for a given ${\bm \alpha}_k$.
Then, we alternatively solve \eqref{eqmmv51} and \eqref{eqmmv52} until the objective function converges.

In the subsequent parts, we develop algorithms to obtain the solutions of problem \eqref{eqmmv51} and problem \eqref{eqmmv52}. Then, we provide the convergence analysis, initialization method, and complexity analysis of the proposed algorithm.

\subsection{Least-Square Solution of ${\bm \alpha}_k $}
In this subsection, we provide the optimal solution of ${\bm{\alpha }}_k$ in \eqref{eqmmv51} when ${\bm {\bar X}}$ is fixed. To begin with, problem \eqref{eqmmv51} is decomposed as the following $K$ individual optimization problems as the variables are decoupled,
\begin{equation}
\begin{aligned}
\mathop {\min }\limits_{{\bm \alpha}_k} \;\left\| {{\bm{\bar Z}}_k^{H} - {\bm{V}}^{H}{{\bm{\alpha }}_k}{{\bm{A}}_R}{{{\bm{\bar X}}}}} \right\|_2^2.
\label{eqmmv521}
\end{aligned}
\end{equation}

Observe that the variable ${\bm{\alpha }}_k$ is a diagonal matrix with $L$ non-zero elements, we vectorize the variables to put the objective function as a function of the $L$ non-zero elements in ${\bm{\alpha }}_k$. Let ${\bm {\bar z}}_k={\rm{vec}}({{\bm{\bar Z}}_k^{H}})\in{\mathbb C}^{BN_f\times1}$  and ${\bm \Upsilon }=({{{\bm{A}}_R}{{{\bm{\bar X}}}}})^T \otimes {{\bm{V}}^H}\in{\mathbb C}^{B{N_f}\times L^2}$.
Note that the $L$ non-zero elements in the vector ${\rm vec}({\bm \alpha}_k)$ are in the following index set ${\Omega _\alpha }$,
\begin{equation}
\begin{aligned}
{\Omega _\alpha }{\rm{ = }}\left\{ {i + (i - 1)L \mid \;{i\; \in \left\{ {1,2,\cdots,L} \right\}} } \right\},
\label{eqmmvindex521}
\end{aligned}
\end{equation}
where $\left| {{\Omega _\alpha }} \right| = L$.
Removing the columns in ${\bm \Upsilon }$ corresponding to the zero entries in ${\rm vec}({\bm \alpha}_k)$, problem \eqref{eqmmv521} is simplified as
\begin{equation}
\begin{aligned}
\mathop {\min }\limits_{{\bm \alpha}_k} \;\left\|  {\bm {\bar z}}_k - {\bm \Upsilon}^{:,\Omega _\alpha}[{\rm vec}({\bm \alpha}_k)]_{\Omega _\alpha } \right\|_2^2.
\label{eqmmv522}
\end{aligned}
\end{equation}
Problem \eqref{eqmmv522} is now a standard LS problem, the optimal solution can be obtained
by equating the first derivative of the objective to zero, which is given by
\begin{equation}
\begin{aligned}
{\bm \alpha}_k={\rm diag}\left(\left(({\bm \Upsilon}^{:,\Omega _\alpha})^H{\bm \Upsilon}^{:,\Omega _\alpha}\right)^{-1}({\bm \Upsilon}^{:,\Omega _\alpha})^H{\bm {\bar z}}_k \right).\label{eqmmv523}
\end{aligned}
\end{equation}
Note that the rank of matrix ${\bm \Upsilon}^{:,\Omega _\alpha}$ in \eqref{eqmmv523} should not be smaller than $L$ for the matrix inverse to exist, which implies that ${\rm rank}({{{\bm{\bar X}}}})\ge \frac{L}{B}$ is required.

\subsection{Iterative Reweighted Algorithm for Optimizing ${{\bm{\bar X}}} $}\label{suboptimizationx}
In this part, we provide an algorithm to find the solution of ${{\bm{\bar X}}} $ in \eqref{eqmmv52} when ${\bm \alpha}_k$'s are fixed.
With fixed ${\bm \alpha}_k$'s , ${\cal L}({{{\bm{\bar X}}}},{\bm \alpha}_k)$ is still a non-convex function, which is hard to solve.
To find an efficient solution of \eqref{eqmmv52}, we apply the iterative reweighted technique to solve it in an iterative manner. The basic principle of this method is to find a family of convex functions which are upper bounds of the objective to approximate the non-convex objective function and solve the approximated problem iteratively. In each iteration, the reweighted coefficients are calculated based on the previously obtained solution and then a new upper bound convex function is formed using the reweighted coefficients. The optimization problem in each
{\mbox{iteration}} is a convex problem, which in our case has a closed-form solution. This technique is also known under different names including, e.g., successive convex
{\mbox{approximation}},
{\mbox{majorization}} minimization and successive upper bound minimization.
 %Note that the approximated function should be the upper bound of the objective function and also should decrease in each iteration to guarantee the convergence.

Denote the value of ${{\bar {\bm X}}}$ at the $t$-th iteration as ${{\bar {\bm X}}(t)}$ and denote ${{\bm{\bar x}}_{i}^H(t)}$ as the $i$-th row vector of constant matrix ${{{\bm{\bar X}}}}(t)$, which are constants in the optimization problem at the $t$-th iteration. Since ${\cal Q}({{\bar {\bm X}}})$ defined in \eqref{eqmmv4} is a concave function in ${{\bar {\bm X}}}$, an upper bound function of ${\cal Q}({{\bar {\bm X}}})$ can be constructed using the first order Taylor approximation at ${{\bar {\bm X}}(t)}$ as
\begin{align}
{\cal Q}({{\bar {\bm X}}}) &= \sum\limits_{i = 1}^{{G_r}} {\log \left( {{\bm{\bar x}}_i^H{{{\bm{\bar x}}}_i} + \varsigma } \right)}\nonumber\\
 &\le \sum\limits_{i = 1}^{{G_r}} {\left( {\frac{{{\bm{\bar x}}_i^H{{{\bm{\bar x}}}_i} + \varsigma }}{{{\bm{\bar x}}_{i}^H(t){{{\bm{\bar x}}}_{i}(t)} + \varsigma }}{\rm{ + }}\log \left( {{\bm{\bar x}}_{i}^H(t){{{\bm{\bar x}}}_{i}(t)} + \varsigma } \right){\rm{ - }}1} \right)}\nonumber\\
 &\buildrel \Delta \over ={\cal Q}^{\rm ub}({{\bar {\bm X}}}|{{\bar {\bm X}}(t)}).\label{eqMM1}
\end{align}
As a result, at each iteration the optimization problem is a weighted least square problem and ${{{\bm{\bar x}}_{i}^H(t){{{\bm{\bar x}}}_{i}(t)} + \varsigma }}$ can be regarded as the inverse of the weighting coefficient at the $t$-th iteration of the objective function, hence the name iterative reweighted algorithm. The inequality can be verified by calculating the first derivation of ${\cal Q}({{\bar {\bm X}}})$, and is omitted here for brevity.  Finally, the upper bound function of the overall objective function ${\cal L}({{\bar {\bm X}}},{\bm \alpha}_k)$ in the $t$-th iteration is given by
\begin{align}
{\cal L}({{{\bm{\bar X}}}},{\bm \alpha}_k)
 &\le {\cal Q}^{\rm ub}({{\bar {\bm X}}}|{{\bar {\bm X}}(t)})+\lambda \sum\limits_{k = 1}^K \left\| {{\bm{\bar Z}}_k^{H} - {\bm{V}}^{H}{{\bm{\alpha }}_k}{{\bm{A}}_R}{{{\bm{\bar X}}}}} \right\|_2^2\nonumber\\
 & \buildrel \Delta \over = {\cal L}^{\rm ub}({{{\bm{\bar X}}}},{\bm \alpha}_k|{{{\bm{\bar X}}}(t)}). \label{eqMM2}
\end{align}
Note that equality holds in \eqref{eqMM2} if ${{{\bm{\bar X}}}}={{\bar {\bm X}}(t)}$. Next, by removing the constant terms in ${\cal Q}^{\rm ub}({{\bar {\bm X}}}|{{\bar {\bm X}}(t)}$ and writing in a matrix form, minimizing ${\cal L}^{\rm ub}({{{\bm{\bar X}}}},{\bm \alpha}_k|{{\bar {\bm X}}(t)})$ is equivalent to solving the following problem
\begin{align}
\mathop {\min }\limits_{{{{\bm{\bar X}}}}} {\rm{Tr}}\left( {{\bm{\bar X}}^H{\bm \Lambda (t)} {{\bm{ \bar X}}}} \right) +\lambda \sum\limits_{k = 1}^K \left\| {{\bm{\bar Z}}_k^{H} - {\bm{V}}^{H}{{\bm{\alpha }}_k}{{\bm{A}}_R}{{{\bm{\bar X}}}}} \right\|_2^2,\label{eqmmvconvex01}
\end{align}
where  ${\bm \Lambda (t)} \in{\mathbb C}^{Gr\times Gr}$ is a diagonal matrix with
\begin{align}
{\bm{\Lambda }(t)} = {\rm{diag}}\left( {\frac{1}{{\bm{\bar x}}_{1}^H(t){{\bm{\bar x}}_{1}(t)} + \varsigma }},\cdots,\frac{1}{{{\bm{\bar x}}_{{G_r}}^H(t){{\bm{\bar x}}_{{G_r}}(t)} + \varsigma }} \right).\label{eqmmvconvex}
\end{align}
It is straightforward to see that problem \eqref{eqmmvconvex01} is convex, and the optimal solution can be obtained by equating the first derivative of the objective function to zero as follows
\begin{align}
{{{\bm{\bar X}(t+1)}}}{\rm{ = }}&{\left( {\frac{{\bm{\Lambda }(t)}}{\lambda } \!+ \!\sum\nolimits_{k = 1}^K {{{( {{\bm{V}}^H{{\bm{\alpha }}_k}{{\bm{A}}_R}})}^H}{\bm{V}}^H{{\bm{\alpha }}_k}{{\bm{A}}_R}} } \!\right)^{\!-1}}\nonumber\\
 &\qquad\qquad\sum\nolimits_{k = 1}^K {{{( {{\bm{V}}^H{{\bm{\alpha }}_k}{{\bm{A}}_R}} )}^H}{\bm{\bar Z}}_k^{H}}.\label{eqooptmalx}
\end{align}

\subsection{Convergence, Initialization, and Complexity Analysis}

\subsubsection{Convergence Analysis}
In this part, we analyze the \mbox{convergence} of the proposed  iterative reweighted based alternative optimization algorithm, which is summarized as Algorithm \ref{algorithm3} with a double loop structure. Specifically, the inner loop is from step 4 to step 6 for optimizing ${\bm {\bar X}}$ and the outer loop is from step 1 to step 9 for alternatively optimizing ${\bm {\bar X}}$ and ${\bm \alpha}_k$. In the following, we show ${\cal L}\left({{{\bm{\bar X}}}^{(r+1)}},{\bm \alpha}_k^{(r+1)}\right)\le{\cal L}\left({{{\bm{\bar X}}}^{(r)}},{\bm \alpha}_k^{(r)}\right)$ to guarantee the convergence of Algorithm \ref{algorithm3} \cite{chen2019resource}, where $r$ is the outer loop iteration index.
Specifically, the proof is given as follows.
\begin{align}
&{\cal L}({\bm{\bar X}}^{(r)},{\bm{\alpha }}_k^{(r)}) \nonumber\\
{\mathop   =  \limits^{({\rm{a}})}}&{\cal L}^{\rm ub}({\bm{\bar X}}^{(r)},{\bm{\alpha }}_k^{(r)}|{{\bm{\bar X}}(0)})
\ge   \mathop {\min }\limits_{{{\bm{\bar X}}}} {\cal L}^{\rm ub}({{\bm{\bar X}}},{\bm{\alpha }}_k^{(r)}|{{\bm{\bar X}}(0)})\nonumber\\
 {\mathop   =  \limits^{({\rm{b}})}}&{\cal L}^{\rm ub}({{\bm{\bar X}}(1)},{\bm{\alpha }}_k^{(r)}|{{\bm{\bar X}}(0)}) \cdots
 \ge    \mathop {\min }\limits_{{{\bm{\bar X}}}} {\cal L}^{\rm ub}({{\bm{\bar X}}},{\bm{\alpha }}_k^{(r)}|{\bm{\bar X}}(t-1))\nonumber\\
 {\mathop   =  \limits^{({\rm{c}})}}&{\cal L}^{\rm ub}({{\bm{\bar X}}(t)},{\bm{\alpha }}_k^{(r)}|{{\bm{\bar X}}(t-1)})
 {\mathop   \ge  \limits^{({\rm{d}})}}  {\cal L}({{\bm{\bar X}}(t)},{\bm{\alpha }}_k^{(r)})\nonumber\\
 {\mathop   =  \limits^{({\rm{e}})}} &{\cal L}({\bm{\bar X}}^{\left( {r + 1} \right)},{\bm{\alpha }}_k^{(r)})
 \ge  \mathop {\min }\limits_{{{\bm{\alpha }}_k}}{\cal L}({\bm{\bar X}}^{\left( {r + 1} \right)},{\bm{\alpha }}_k^{(r)})\nonumber\\
 {\mathop   =  \limits^{({\rm{f}})}}&  {\cal L}({\bm{\bar X}}^{\left( {r + 1} \right)},{\bm{\alpha }}_k^{(r + 1)}),
\end{align}
where (\rm a) holds because we initialize ${\bm{\bar X}}^{(r)}={\bm{\bar X}}(0)$ and the equality holds in \eqref{eqMM2} if and only if ${\bm{\bar X}}^{(r)}={\bm{\bar X}}(0)$. In (\rm b) and (\rm c) we use the fact that ${\bm{\bar X}}(t)$ is the optimal solution of \eqref{eqmmvconvex01}. In (\rm d) we use \eqref{eqMM2}; (\rm e) holds as we update ${\bm{\bar X}}^{(r+1)}$ as ${\bm{\bar X}}(t)$; and in (\rm f)  ${\bm{\alpha}}_{k}$ is the optimal solution of \eqref{eqmmv521}.

Therefore, we have ${\cal L}\left({{{\bm{\bar X}}}^{(r+1)}},{\bm \alpha}_k^{(r+1)}\right)\le{\cal L}\left({{{\bm{\bar X}}}^{(r)}},{\bm \alpha}_k^{(r)}\right), ~\forall r$, this implies that the objective function is montonically decreasing with the number of iterations. Together with the fact that the objective function is bounded below guarantees the convergence of Algorithm \ref{algorithm3}.

\subsubsection{Initialization Analysis}
It is worth noting that a proper initialization in  Algorithm \ref{algorithm3} is crucial for the successful recovery of the sparse channel matrix, especially for the initialization of ${\bm \alpha}_k$.
This is because only when the initialized ${\bm \alpha}_k$ is as close as possible to its actual value, all the cascaded channels have the similar sparse channel matrices that can be jointly recovered due to \eqref{eqscaling1}. Otherwise, there is no solution of ${{\bar {\bm X}}}$ that makes the equality in \eqref{eqscaling1} hold.

In fact, we can initialize ${\bm \alpha}_k$ by first estimating the cascaded channels ${\bm G}_k$ individually, i.e., using the SMV method, using the MMV method, or solving the single user version of \eqref{eqmmv3}. Then, we use the estimated ${\hat {\bm G}}_k$  to initialize ${\bm \alpha}_k$ according to the scaling property \eqref{eqscaling} with
\begin{equation}
\begin{aligned}
{{\bm{\alpha }}_k^{l,l}} = \frac{{{{[ {{\bm{h}}_{{k}}^H} ]}_l}}}{{{{[ {{\bm{h}}_{{1}}^H}]}_l}}}\approx
\frac{{\hat{\bm{G}}_{k}^{l,1}}}{{\hat{\bm{G}}_{1}^{l,1}}}\approx\frac{{\hat{\bm{G}}_{k}^{l,2}}}{{\hat{\bm{G}}_{1}^{l,2}}} \cdots \approx\frac{{\hat{\bm{G}}_{k}^{l,M}}}{{\hat{\bm{G}}_{1}^{l,M}}},\; \forall k, l\label{eqscalingex0}.
\end{aligned}
\end{equation}
Thus, we can initialize ${\bm \alpha}_k^{(0)}$ as
\begin{align}
\left({\bm \alpha}_k^{(0)}\right)^{i,l}= \left\{ {\begin{array}{*{20}{l}}
\frac{1}{M}\sum\nolimits_{m = 1}^M \frac{{\hat{\bm{G}}_{k}^{l,m}}}{{\hat{\bm{G}}_1^{l,m}}},\; {\rm if} \;1\le i=l\le L,
\\
0,\quad\quad\quad\quad\quad\quad\;{\rm otherwise }.\label{eqinitialization}
\end{array}} \right.
\end{align}
Note that given the cardinality of ${\bm G}_k$, estimating cascaded channels ${\bm G}_k$ requires at least $2N_fN_{h_k}$ measurements to form a set of $2 N_f N_{h_k}$ equations with $2N_fN_{h_k}$ unknown variables (the indices and values of the nonzero entries in each {\mbox{column}} of its sparse matrix) \cite{elad2007optimized,foucart2017mathematical}. In practice, it requires $cN_fN_{h_k}\ln(\frac{eL}{N_fN_{h_k}})$ measurements for a stable solution, where $c>0$ is a constant depending on the stability requirement \cite{foucart2017mathematical}. Combining with the requirements in \eqref{eqmmv523}, the minimum training overhead $B$  should satisfy the following constraint: $B \ge \max\left(cN_fN_{h_k}\ln(\frac{eL}{N_fN_{h_k}}),\frac{L}{{\text {rank}}( \bm{ \bar X})}\right) $.

\begin{algorithm}[t]{\small{
\caption{Iterative Reweighted based Alternative Optimization}\label{algorithm3}
%\caption{BCD associated with CCCP for $({\rm {\bm P}2{\rm}})$}\label{algorithmP0}
\begin{algorithmic}[1]
 \STATE Initialize ${\bm {\bar X}}^{(0)}$ as ${\bm 1}_{G_r\times {\hat {N_f}}}$, ${\bm \alpha}_k^{(0)}$ as \eqref{eqinitialization}, and the outer loop iteration counter $r=0$.
 \REPEAT
 \STATE Set ${\bm \alpha}_k={\bm \alpha}_k^{(r)}$ and ${\bm {\bar X}}(0)={\bm {\bar X}}^{(r)}$, initialize the inner loop iteration counter $t=0$.
 \REPEAT
 \STATE Given ${\bm {\bar X}}(t)$, calculate ${\bm {\bar X}}(t+1)$ based on \eqref{eqooptmalx}, and then set $t= t+1$.
 \UNTIL ${\cal L}^{\rm ub}({{{\bm{\bar X}}}},{\bm \alpha}_k|{{{\bm{\bar X}}}(t)})$ converges.
  \STATE Set $r=r+1$ and update $ {\bm {\bar X}}^{(r)}={\bm {\bar X}}(t)$, then given $ {\bm {\bar X}}= {\bm {\bar X}}^{(r)}$, calculate ${\bm \alpha}_k^{(r)}$ based on \eqref{eqmmv523}.
 \UNTIL  ${\cal L}({{{\bm{\bar X}}}^{(r)}},{\bm \alpha}_k^{(r)})$ converges.
\end{algorithmic}}}
\end{algorithm}

\subsubsection{Complexity Analysis}
To characterize the computational complexity, we count the number of complex multiplication (CM) \cite{chen2019Exploiting} in Algorithm 3.  The main complexity comes from computing \eqref{eqooptmalx} in Step 5 in the inner loop, and computing \eqref{eqmmv523} in Step 7 in the outer loop. It is straightforward to see that the two operations require ${\Delta _1}{\rm{ = }}{{{G}}_r}\left( {N_f + 1} \right) + K\left( {BL\left( {1 + {{\rm{G}}_r}} \right) + G_r^2B + B{G_r}{N_f}} \right) + \left( {{N_f}{\rm{ + }}\beta } \right)G_r^2$ CMs and ${\Delta _2}{\rm{ = }}K\left( {2{L^2}B{N_f} + \beta {L^2} + LBN_f} \right)$ CMs, respectively. Then, if we denote $I_{\rm in}$ and $I_{\rm out}$ as the number of iterations required for the inner loop and the outer loop, respectively, the total computational complexity can be expressed as ${\cal O}\left( I_{\rm in}I_{\rm out}{{\Delta _1}{\Delta _2}} \right)$.
\section{ Training Reflection Coefficients Optimization}\label{sec:Pilotoptimization}
In this section, the optimization of the training sequence of reflection coefficients at
the RIS is studied.

To optimize the reflection coefficients in the training phase, we follow the principle from \cite{elad2007optimized} that a better recovery performance can be achieved if the mutual coherence of the equivalent dictionary is smaller.
As a heuristic, from \eqref{eq14}, we see that ${\bm{D}}={\bm V}^{H}{\bm A}_R$ can be regarded as the equivalent dictionary. In this case, the corresponding mutual coherence is given by
\begin{equation}
\begin{aligned}
 \mu \left( {\bm{D}} \right){\rm{ = }}\mathop {\max }\limits_{i \ne j,1 \le i,j \le {G_r}} \left\{ \frac{{{{({\bm{D}}^{:,i})}^H{{\bm{D}}^{:,j}}}}}{{{{\left\| {{{\bm{D}}^{:,i}}} \right\|}_2}{{\left\| {{{\bm{D}}^{:,j}}} \right\|}_2}}} \right\}.
 \end{aligned}
\end{equation}
Minimizing the mutual coherence implies that the columns of $\bm D$ should be as orthogonal as possible. Equivalently, it requires designing $\bm V$ to make ${\bm D}^{H}{\bm D}$ as close as possible to a scaled identity matrix, i.e.,
\begin{equation}
\begin{aligned}
{\bm{D}}^{H}{\bm{D}}=  {{\bm{A}}_R^{H}{\bm{V}}{{\bm{V}}^{H}}{{\bm{A}}_R}}  \approx B{{\bm{I}}_{{G_r}}},\label{eqidentity}
\end{aligned}
\end{equation}
where $B$ is a constant for normalization. Note that ${\bm A}_R$ in \eqref{eqidentity} is a constant dictionary given by the quantization of the angles, which is defined in \eqref{ARdefine}.

The solution for problem \eqref{eqidentity} for unconstrained $\bm{V}$ has been investigated in \cite{duarte2009learning}.
However, since RIS just induces phase shifts on the incident signals without changing their \mbox{amplitudes}, the reflection coefficients should satisfy the \mbox{following} constraint \cite{abeywickrama2019intelligent,wu2018beamformingOptimization}, i.e.,
\begin{eqnarray}
\left| {{\bm{V}}^{l,b}} \right| = 1, \forall l,b.\label{eqconstraint}
\end{eqnarray}
Thus, the method in \cite{duarte2009learning} cannot be directly applied to find the solution of $\bm V$ in \eqref{eqidentity}.

In the following, we modify the method in \cite{duarte2009learning} to solve problem \eqref{eqidentity} subject to constraint \eqref{eqconstraint}. First, we multiply $\bm{A}_R$ and $\bm{A}_R^H$ on both sides of \eqref{eqidentity} and get the following equation:
\begin{eqnarray}
{{\bm{A}}_R} {\bm{A}}_R^{H}{\bm{V}}{{\bm{V}}^{H}}{{\bm{A}}_R}{{\bm{A}}_R^{H}}  \approx B{{\bm{A}}_R}{{\bm{A}}_R^{H}}.\label{eqV2}
\end{eqnarray}
Let ${{\bm{U}}_R}{\bm \Xi}{\bm{U}}_R^{H}$ be the eigenvalue decomposition of ${\bm A}_R{\bm A}_R^{H}$, where ${\bm \Xi }={\rm diag}({\gamma _1},{\gamma _2},\cdots,{\gamma _L})\in{\mathbb C}^{L\times L}$ is the eigenvalue matrix with the eigenvalues $\gamma_l$ ordered in a decreasing order in magnitude.
Then, substituting ${{\bm{U}}_R}{\bm \Xi}{\bm{U}}_R^{H}={{\bm{A}}_R}{{\bm{A}}_R^{H}}$ and ${\bm{U}}_R{\bm{U}}_R^{H}={{\bm{U}}_R}^{H}
{\bm{U}}_R={\bm I}_{L}$ into \eqref{eqV2}, we have the following reformulated reflection coefficients optimization problem, i.e.,
\begin{equation}
\begin{aligned}
\!\!\!\!\mathop {\min }\limits_{\bm{V}} \left\| B{\bm \Xi} -{ \bm \Xi}{\bm{U}}_R^{H}{\bm{V}}{{\bm{V}}^{H}}{{\bm{U}}_R}{\bm \Xi}  \right\|_2^2 \;{\rm{s}}{\rm{.t}}{\rm{.}}\;\left| {{\bm{V}}^{l,b}} \right| = 1, \forall l,b. \label{eqV3}
%\left| {{\bm{V}}\left( {b,l} \right)} \right| = 1, \forall b,\forall l, \label{eqV3}%1\le b\le B,1\le l\le L.
\end{aligned}
\end{equation}

Let ${{\bm Q}}={\bm \Xi} {\tilde {\bm Q}}={\bm \Xi}{\bm{U}}_R^{H}{\bm{V}}$, where ${\tilde {\bm Q}}={\bm{U}}_R^{H}{\bm{V}}$.
Denoting the $b$-th column vector of $\bm Q$, $\bm {\tilde Q}$, and  $\bm V$ as ${\bm q}_b$, ${\bm {\tilde q}}_b$, and  ${\bm v}_b$, respectively, we have
\begin{equation}
\begin{aligned}
{{{\bm{ q}}}_b} ={\bm \Xi} {{\tilde{\bm{ q}}}_b} ={[{{{\gamma _1}}}{{{{\tilde q}_{b,1}}}},{{{\gamma _2}}}{{{{\tilde q}_{b,2}}}},\cdots,{{{\gamma _L}}}{\tilde q}_{b,L}]^T}\label{eq65},
%\\{{\bf{q}}_b} = {\bf{\Xi }}{{{\bf{\tilde q}}}_b}\\
%{{{\bf{\tilde q}}}_b} = {\bf{U}}_R^H{{\bf{v}}_b}
\end{aligned}
\end{equation}
where ${{{\bm{\tilde q}}}_b} = {\bm{U}}_R^H{{\bm{v}}_b} \buildrel \Delta \over ={[{\tilde q_{b,1}},{\tilde q_{b,2}}, \cdots ,{\tilde q_{b,L}}]^T} \in {{\mathbb C}^{L \times 1}} $.

Next, we will first optimize ${\bm q}_b$ successively and use the \mbox{optimized} ${\bm q}_b$ to obtain the solution of ${\bm v}_b$ heuristically.
\mbox{Specifically},  since ${\bm{Q}}{{\bm{Q}}^H} = \sum\nolimits_{i = 1}^B {{{\bm{q}}_i}{\bm{q}}_i^H} $, the optimization of ${\bm v}_b$ in \eqref{eqV3} can be rewritten as
\begin{equation}
\begin{aligned}
\mathop {\min }\limits_{{\bm{v}}_b} \left\| {
 {B{\bm \Xi} - \sum\nolimits_{i \ne b}^B {{{\bm{q}}_i}{\bm{q}}_i^{H}}}{\rm{ - }}{{\bm{q}}_b}{\bm{q}}_b^{H}} \right\|_2^2\;\;\;{\rm{s}}{\rm{.t}}{\rm{.}}\;\left| {{\bm{V}}^{l,b}} \right| = 1, \forall l.\label{eqV4}
\end{aligned}
\end{equation}
where ${{\bm{E}}_b} = B{\bm \Xi} - \sum\nolimits_{i \ne b}^B {{{\bm{q}}_i}{\bm{q}}_i^{H}}\in{\mathbb C}^{L\times L} $ is regarded as a constant because ${\bm q}_i$ for $i\neq b$ is fixed when optimizing ${\bm q}_b$. Note that ${\bm q}_b$ is a function with respect to ${\bm v}_b$ in problem \eqref{eqV4}.

Let ${\bar{\bm{U}}_{b}}{{\bm\Psi} _{b}}{\bar{\bm{U}}}_{b}^{H}=\sum\nolimits_{l = 1}^L {{\xi _{b,l}}{{{\bm{\bar u}}}_{b,l}}{\bm{\bar u}}_{b,l}^H} $ be the \mbox{eigenvalue} \mbox{decomposition} of ${\bm E}_b$, and  ${{\bm\Psi} _{b}}={\rm diag}({\xi  _{b,1}},{\xi  _{b,2}},\cdots,{\xi  _{b,L}})$ $\in{\mathbb C}^{L\times L}$ be the eigenvalue matrix where the eigenvalues $\xi_{b,l}$ are ordered in a decreasing order in magnitude.
Then, given ${\bm E}_b$ in \eqref{eqV4}, the solution of ${\bm q}_b$ is
\begin{equation}
\begin{aligned}
{\bm q}_b=\sqrt{\xi_{b,1}}{\bar{\bm u}}_{b,1}\label{eqeq157},
\end{aligned}
\end{equation}
where ${\bar {\bm u}}_{b,1}$ is the first column vector of ${{\bar {\bm{U}}}_{b}}$ corresponding to the largest eigenvalue of ${\bm E}_b$. This solution can eliminate the largest error in \eqref{eqV4} for a given ${\bm E}_b$ .

Then, since ${{{\bm{ q}}}_b} ={\bm \Xi} {{\tilde{\bm{ q}}}_b}$ as defined in \eqref{eq65}, we can recover ${\bm {\tilde q}}_b$ from the optimized ${{{\bm{ q}}}_b}$ in \eqref{eqeq157}\footnote{Note that since matrix ${\bm E}_b$ is not full-rank in general, we only need to update the components in ${\bm {\tilde q}}_b$ corresponding to the positive eigenvalues.}.
Furthermore, since ${{{\bm{\tilde q}}}_b} = {\bm{U}}_R^H{{\bm{v}}_b} $ and \eqref{eqconstraint}, we need to solve the following projection problem to obtain the solution of ${\bm v}_b$:
\begin{equation}
\begin{aligned}
\mathop {\min }\limits_{{{\bm{v}}_b}} \left\| {\bm{U}}_R^H{{{\bm{v}}_b}{\rm{ - }}{{{\bm{\tilde q}}}_b}} \right\|_2^2\;\;\;\;{\rm{s}}{\rm{.t}}{\rm{.}}\;\left| {\bm{V}}^{l,b}  \right| = 1, \forall l. \label{eqproj}
\end{aligned}
\end{equation}
From \cite{chen2019intelligent},  the optimal solution of \eqref{eqproj} is given by $
{{\bm{v}}_b}={{e}}^{{{j}}\angle {({\bm U}_R{\bm{\tilde q}}_b)}}.$
Then, we substitute ${{\bm{v}}_b}$ into \eqref{eqV3} and repeat the above steps $B$ times to compute ${{\bm{v}}_1},{{\bm{v}}_2},\cdots,{{\bm{v}}_B}$,  successively.
\mbox{Finally}, we can obtain the optimized training reflection \mbox{coefficients} ${\bm V}$. %Note that the above proposed reflection \mbox{coefficients} \mbox{optimization}  is not an \mbox{optimal} algorithm but a non-iterative heuristical algorithm. %The steps of the algorithm are  \mbox{summarized} as Algorithm \ref{algorithm41}.

\section{Simulation Results}\label{sec:Simulation}

%In this section, we  show the simulation results to validate the effectiveness of the proposed algorithm.
In all simulations, we assume $\alpha_p$ and $\beta_{k,q}$ follow complex Gaussian distribution with unit power. $\phi _p^{\rm AoA}$, $\phi _p^{\rm AoD}$, and ${\varphi_{k,q}}$ are continuous and uniformly distributed over $[-\pi/2,\pi/2)$. Parameters $\varsigma$, $d$, $G_r$, $G_t$, and $N_{h_k}$ are set as $10^{-9}$, 0.1, 512, 128, and 1, respectively.
Set $\varpi/\rho=1/2$ where $\varpi $ is the antenna spacing and $\rho $ is the carrier wavelength.
In addition, the noise power ${\delta^2}$ is normalized to one, and the transmit power $P$ (dB) is described as a relative value of the noise power. Specifically, we use the normalized mean square error (NMSE) as the performance metric \cite{tsai2018efficient}, which is given by
\begin{align}
{\rm NMSE}={\mathbb E}\left[{{\left\| {{{\bm{\hat G}}_k} - {\bm{G}}_k} \right\|_2^2} \mathord{\left/
 {\vphantom {{\left\| {{{\bm{\hat  G}}_k} - {\bm{G}}_k} \right\|_2^2} {\left\| {\bm{G}} \right\|_2^2}}} \right.
 \kern-\nulldelimiterspace} {\left\| {\bm{ G}}_k \right\|_2^2}}\right].
\end{align}

In the simulations, we compare the NMSE performances obtained from the following channel estimation schemes. Results for each scheme are averaged over $500$ Monte Carlo trials.
\begin{itemize}
\item {\emph {LS}}: The channels are estimated using the estimation protocol in Fig. \ref{Fig2Frame} and the LS estimator in \eqref{eqLSestimator} with the optimal training sequences studied in \cite{jensen2019optimal}.%, which requires $B\ge L$;
\item  {\emph {Binary Reflection}} \cite{mishra2019channel}: The channels are estimated by turning on only one reflective element and keeping the rest reflective elements closed.%,  which also requires $B\ge L$;
%\item {\emph {Genie-aided LS}}: The cascaded channels are estimated by assuming that the BS knows the exact angles of the cascaded AoAs at the RIS/user side and the AoDs at the BS side, and using the LS estimator.
    %to estimate ${\bm x}_k$, and estimating the cascaded channel by \eqref{eqchannel}, which is the performance upper bound.
\item {\emph {MMV}}: The channels are estimated by formulating a MMV problem, and solved using SOMP algorithm as shown in Algorithm \ref{algorithm2}.
\item {\emph {S-MMV}}: The channels are estimated by projecting the received signals onto the common subspace first, and solved using the MMV estimator.
\item {\emph  {S-SMV}}: The channels are estimated by projecting the received signals onto the common subspace first, and solved using a SMV problem  with the OMP algorithm \cite{tsai2018efficient}.
%\item {\emph {S-Individual iterative reweighted}} (IR): The channels are estimated by projecting the received signals into the common subspace firstly, and formulating a MMV problem similar as \eqref{eqsmv} for a individual single-user, and just applying Algorithm \ref{algorithm3} with initializing ${\bm \alpha}_k={\rm diag}\left([1,1,\cdots,1]\right)\in{\mathbb C}^{L\times L}$.
\item {\emph {S-MJCE}}: The channels are estimated by the proposed two-step (subspace) multi-user joint channel estimation procedure, i.e., Algorithm \ref{algorithmSub}.
\item {\emph {S-Genie-aided LS}}:  The cascaded channels are estimated by assuming that the BS knows the exact angles of the cascaded AoAs at the RIS/user side and the AoDs at the BS side, and projecting the received signals onto the known common AoD subspace with the solution of LS estimator. This scheme serves as a performance lower bound for all other schemes.
\end{itemize}
%
% \begin{figure}
%  \begin{minipage}{.48\textwidth}
%   \centering
%   \includegraphics[scale=0.5]{figure_overhead1.eps}
%\caption{Effect of training overhead $B$ on the NMSE: $P=10$dB, $K=T=4$, $L=128$, and $N_f=8$.}
% \label{figureoverhead1}
% \end{minipage}\quad
% \begin{minipage}{.48\textwidth}
%   \centering
%   \includegraphics[scale=0.5]{figure_sparsity.eps}
%\caption{Effect of the number of scatterers between the BS and the RIS on the NMSE: $P=10$dB, $B=30$, $K=T=4$, and $M=L=128$.} \label{figuresparsity}
%  \end{minipage}
% \end{figure}

\begin{figure}[t]
\centering
\includegraphics[scale=0.6]{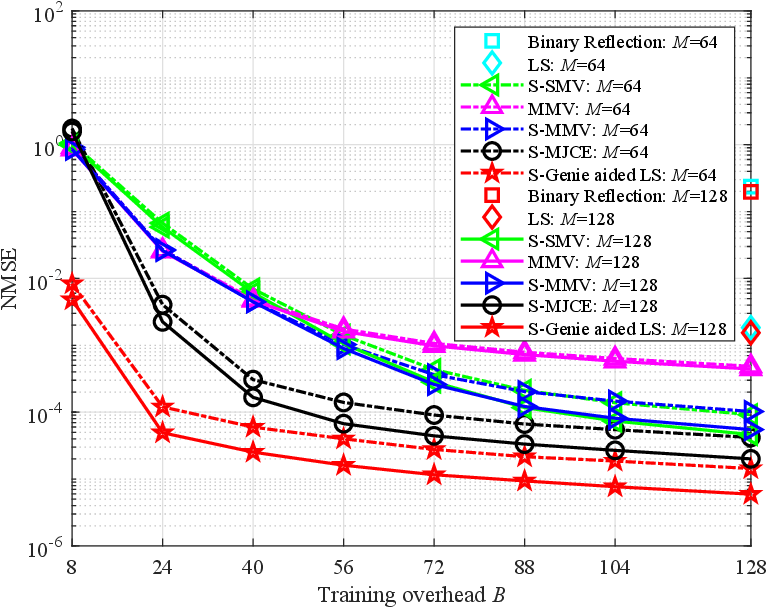}
\caption{Effect of training overhead $B$ on the NMSE: $P=10$dB, $K=T=4$, $L=128$, and $N_f=8$.}
 \label{figureoverhead1}
\end{figure}

\begin{figure}[t]
\centering
\includegraphics[scale=0.6]{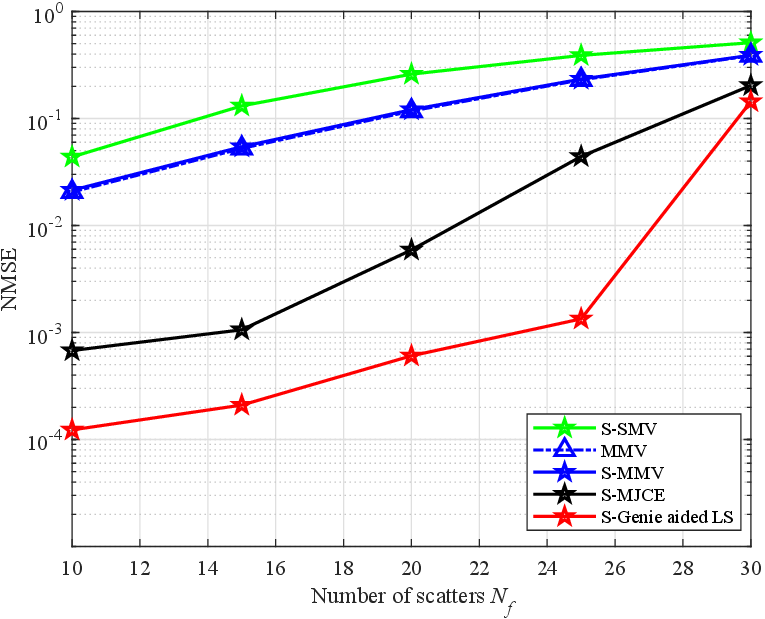}
\caption{Effect of the number of scatterers between the BS and the RIS on the NMSE: $P=10$dB, $B=30$, $K=T=4$, and $M=L=128$.} \label{figuresparsity}
\end{figure}
\begin{figure}[t]
\centering
 \includegraphics[scale=0.6]{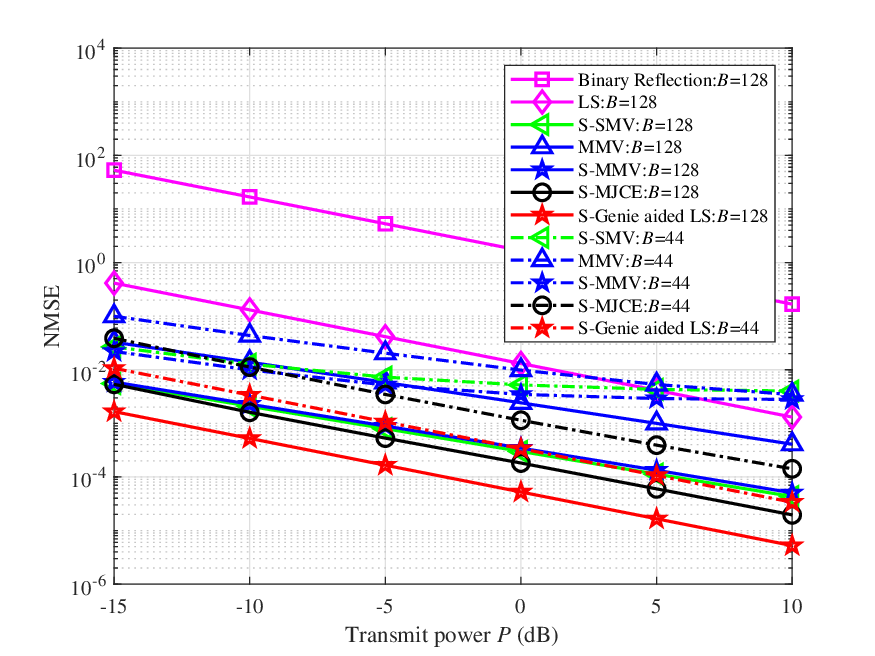}
\caption{Effect of the transmit power on the NMSE: $K=T=4$, $M=L=128$, and $N_f=8$.}
\label{figureSINR}
\end{figure}

%\begin{figure}[t]
%\centering
%\begin{minipage}[t]{0.48\textwidth}
%\centering
% \includegraphics[scale=0.48]{figure_overhead1.eps}
%\caption{Effect of training overhead $B$ on the NMSE: $P=10$dB, $K=T=4$, $L=128$, and $N_f=8$.}
%\label{figureoverhead1}
%\end{minipage}\quad
%\begin{minipage}[t]{0.48\textwidth}
%\centering
%\includegraphics[scale=0.48]{figure_sparsity.eps}
%\caption{Effect of the number of scatterers between the BS and the RIS on the NMSE: $P=10$dB, $B=30$, $K=T=4$, and $M=L=128$.}
%\label{figuresparsity}
%\end{minipage}
%\end{figure}

Fig. \ref{figureoverhead1} shows the impact of training overhead $B$ on the NMSE.
First, we observe that the estimation performances of all schemes improve as the training overhead $B$ increases.
%This is because increasing the number of measurements can reduce the noise perturbation, thus achieving a lower noise variance, and further achieving a better successful recovery and estimation accuracy.
Next, the performance of the binary reflection method is worse than that of other baselines. This is because in each time slot, the power of training reflection coefficient at the RIS of this method is one while that of other schemes is $L$.
We observe that the performance gap between the MMV and the S-MMV increase with $B$, which implies that subspace projection can improve the estimation performance significantly with a large $B$ for the studied system.
Finally, the proposed method outperforms other baseline schemes significantly, and achieves a similar performance as the genie-aided lower bound. This validates the effectiveness of the proposed estimator.

Fig. \ref{figuresparsity} shows the impact of the number of scatterers between the BS and the RIS on the NMSE.
First, we observe that the performances of all estimation schemes decrease as the number of \mbox{scatterers} grows, especially when the number of \mbox{scatterers} equals the training overhead $B$. This is because the number of unknown parameters required to be estimated increases with the number of scatterers.
 Also, the performance gaps among these methods decrease as the number of \mbox{scatterers} grows. This is because the estimation performance is limited by the number of measurements and SNR. Besides, the proposed method outperforms other baseline schemes especially when the number of scatterers is small, which also validates the effectiveness of the proposed estimator.

Fig. \ref{figureSINR} shows the impact of transmit power on the NMSE. We observe that the performances of all schemes improve with the transmit power. This is because a higher  SNR means less noise perturbation, thus a better recovery/estimation performance.
Also, the slopes of the NMSE curves of S-SMV/MMV/S-MMV decrease with transmit power. This is because these schemes are based on matching pursuit algorithm, which are limited by the angular resolution of the dictionary matrix.
Moreover, comparing with other benchmarks, we observe that the proposed algorithm has relatively poor  performance at smaller training overhead  or lower SNR but better performance at larger training overhead and higher SNR. This is because smaller training overhead and lower SNR cause imprecise estimation of ${\bm \alpha}_k$ and improper initialization of the proposed algorithm.

\begin{figure}[t]
\centering
\includegraphics[scale=0.6]{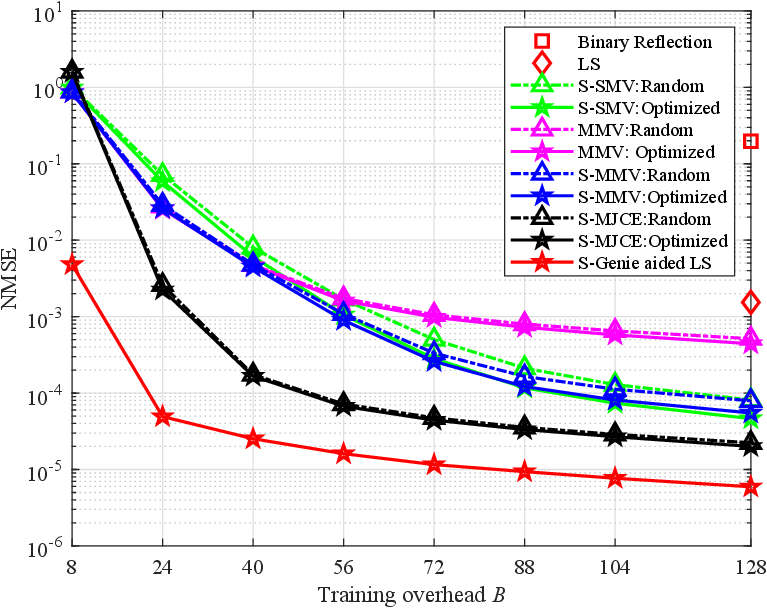}
\caption{Effect of training reflection coefficient optimization on the NMSE: $P=10$dB, $K=T=4$, $M=L=128$, and $N_f=8$.}\label{figurepilotdesign}
\end{figure}

% \begin{figure}
%  \begin{minipage}{.48\textwidth}
%   \centering
%  \includegraphics[scale=0.45]{figure_SINR1.eps}
%\caption{Effect of the transmit power on the NMSE: $K=T=4$, $M=L=128$, and $N_f=8$.}
%\label{figureSINR}
% \end{minipage}\quad
% \begin{minipage}{.48\textwidth}
%   \centering
%  \includegraphics[scale=0.45]{figure_pilotdesign.eps}
%\caption{Effect of training reflection coefficient optimization on the NMSE: $P=10$dB, $K=T=4$, $M=L=128$, and $N_f=8$.}\label{figurepilotdesign}
%  \end{minipage}
% \end{figure}

Fig. \ref{figurepilotdesign} shows the impact of  designing the training reflection coefficients on the NMSE. We observe that the performance gaps for all estimators using the optimized and randomly generated training reflection coefficients increase as the \mbox{pilot} length increases.
This is because the dimension of the \mbox{equivalent} dictionary increases with the pilot length. Columns of the equivalent dictionary with a higher dimension can be more orthogonal after the optimization, thus achieving a smaller mutual coherence. Based on the principle from \cite{elad2007optimized}, a better recovery performance can be achieved if the mutual coherence of the equivalent dictionary is smaller.
%However, compared with the S-MMV and MMV based estimator, the performance gap of the proposed estimator with the optimized and randomly generated training reflection coefficients is very small. This is because we apply all the received signals to recover the sparse matrix jointly, which reduces the impact of  designing the training reflection coefficients.

\begin{figure}[t]
\centering
\includegraphics[scale=0.6]{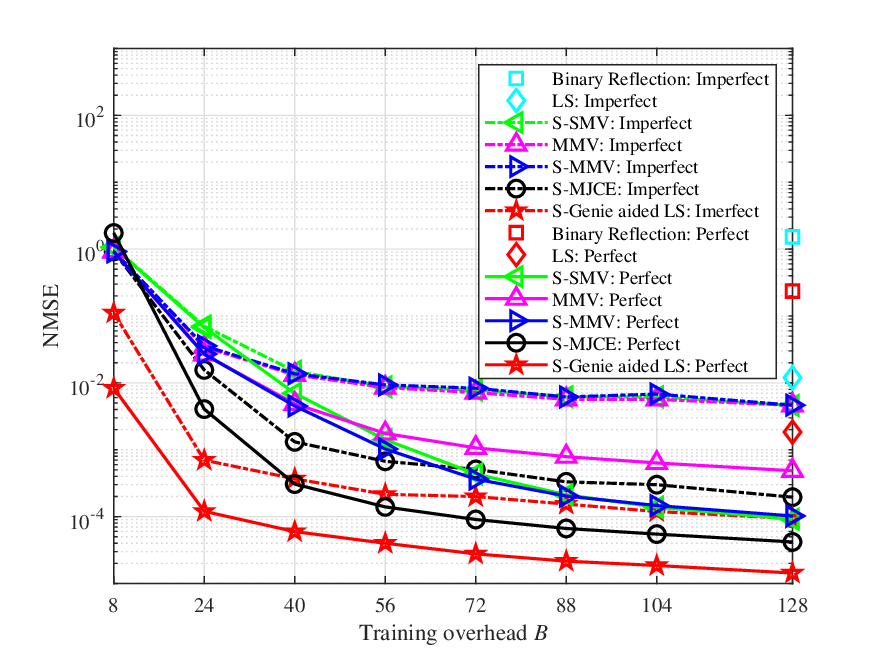}
\caption{Effect of direct link channel estimation error on the NMSE: $K=T=4$, $M=L=128$, $P=10$dB, and $N_f=8$.}
\label{figureSINR2}
\end{figure}

\begin{figure}[t]
\centering
\includegraphics[scale=0.6]{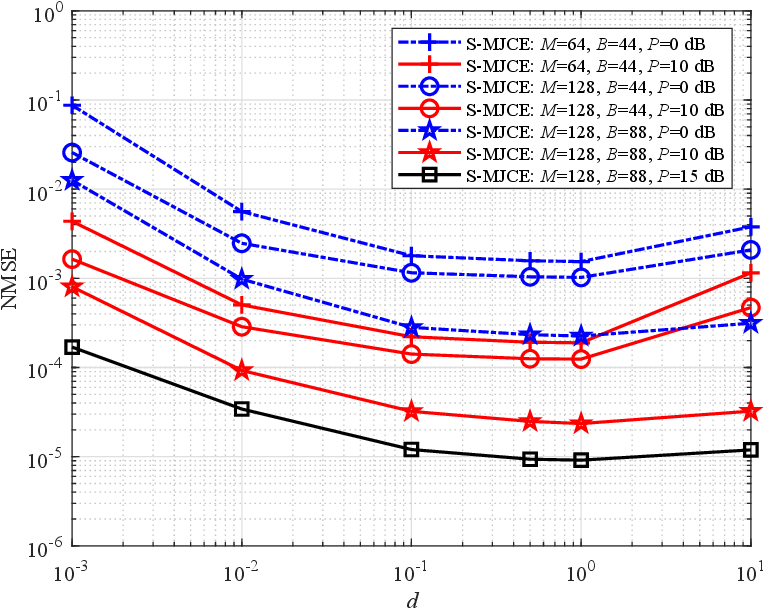}
\caption{Effect of $\lambda={\textstyle \frac{{PT}d}{\delta^2{\rm log}{G_r}}}$ on the NMSE: $K=T=4$, $L=128$, and $N_f=8$.}
\label{figurelambda1}
\end{figure}

 \begin{figure}[t]
\centering
\includegraphics[scale=0.6]{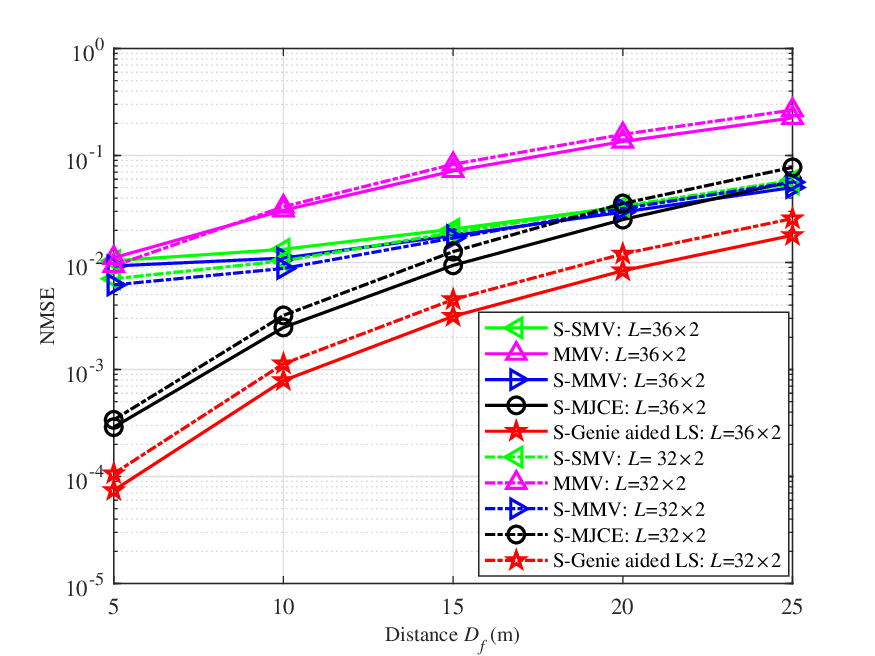}
\caption{Effect of the BS-RIS distance on the NMSE:  $K=T=4$, $B=35$, $M=128$, $G_r=128\times4$, and $N_f=6$.}
\label{figureSINR3}
\end{figure}
%
%
%\begin{figure}[t]
%\centering
%\begin{minipage}[t]{0.48\textwidth}
%\centering
%\includegraphics[scale=0.45]{figureerror.eps}
%\caption{Effect of direct link channel estimation error on the NMSE: $K=T=4$, $M=L=128$, $P=10$dB, and $N_f=8$.}
%\label{figureSINR2}
%\end{minipage}
%\quad
%\begin{minipage}[t]{0.48\textwidth}
%\centering
% \includegraphics[scale=0.45]{figure_lambdaselection.eps}
%\caption{Effect of $\lambda={\textstyle \frac{{PT}d}{{\rm log}{G_r}}}$ on the NMSE: $K=T=4$, $L=128$, and $N_f=8$.}
%\label{figurelambda1}
%\end{minipage}
%\end{figure}

Fig. \ref{figureSINR2} shows the NMSE performances under perfect CSI and imperfect CSI of the direct link channel, where the direct link channel estimation error is given in \eqref{eqsignal1}. We observe that the performances of all schemes under imperfect CSI are worse than those under perfect CSI.
 This is because the direct link channel estimation error adds to the signal model as extra noise, which  leads to a lower SNR.
The performance gaps between MMV and S-SMV/S-MMV under perfect CSI are larger than those under imperfect CSI, which means that the performance of the subspace projection degrades under imperfect CSI due to the residual estimation errors.
However, the performance gap between the proposed algorithm and the lower bound benchmark is not affected by the imperfect CSI, which validates the effectiveness of the proposed algorithm.

Fig. \ref{figurelambda1} shows the impact of penalty factor $\lambda$ on the NMSE. We observe that the performance of the proposed estimator decreases first and then increases as $\lambda$ grows.
This is because the penalty factor is applied for balancing the tradeoff between data fitting and the sparsity of the solutions.
Hence, a smaller $\lambda$ causes a more sparse matrix but does not fit the data, while a larger $\lambda$ fits the data well but loses the sparsity. Both of the two scenarios degrade the recovery performance of the proposed estimator.

Fig. \ref{figureSINR3} shows the effect of path-loss and RIS size with UPA on the NMSE performance.
We use the parameters given in Table XIV of \cite{hemadeh2017millimeter} to model the path-loss in mmWave \mbox{communications}.
Specifically, we consider a carrier frequency of 28GHz with radio frequency band 400MHz from 5G new radio standards in 3GPP Release 15\cite{3GppTR15}. The path-loss is  $\left(61.4+34.1\log_{10} D/D_0\right)$ dB \cite{hemadeh2017millimeter}, where $D$ is the \mbox{distance} and $D_0=1$m is the reference distance.
The noise power spectral density, the transmit power, transmit antenna gain, and received antenna gain are set as  -174dBm/Hz, 40dBm, 17dBi, and 25dBi, respectively.
In addition, the \mbox{distances} from the BS to  the RIS and from  the RIS to $U_k$ are denoted by $D_f$ and $D_k=3$m, respectively.
We observe that the performances decrease with $D_f$, because a larger distance lowers the SNR. Further, the proposed algorithm has worse performance than other schemes when $D_f\ge20$m. This is \mbox{because} lower SNR gives a worse initialization for the \mbox{proposed} algorithm.
In addition, the performances increase when the RIS size changes from $32\times2$ to $36\times2$. This is because the number of unknown variables remains unchanged and a better mutual coherence and a higher SNR may be achieved. Finally, the results confirm that our proposed \mbox{algorithm} works well in practical setups.

\section{Conclusions}\label{sec:Conclusion}
This paper studies the channel estimation problem for the RIS aided multi-user mmWave MIMO system and \mbox{proposes} a novel channel estimation protocol to estimate the \mbox{cascaded} channels directly. Using the observation that the cascaded channels are typically sparse, we formulate the channel \mbox{estimation} problem as a sparse channel matrix \mbox{recovery} \mbox{problem} using CS techniques, which enables \mbox{channel} \mbox{estimation} with limited training overhead.
In particular, the sparse channel matrices of the cascaded channels of all users have a common sparsity structure due to the common channel between BS and RIS. Hence, by taking advantage  of such a common sparsity structure, we propose a two-step multi-user joint \mbox{channel} \mbox{estimation} procedure.
Moreover, the \mbox{optimization} of the \mbox{training} sequence of reflection coefficients at the RIS is studied.
Finally, the simulation results validate the \mbox{effectiveness} of the proposed estimator.

\appendix{}

\section{}\label{appendixvA}

In this appendix, we first derive the likelihood function associated with the AoD vectors in ${\bm{A}}_T^{:,{\Omega _{\rm{D}}}}$, and then present the solution for subspace estimation using the maximum likelihood approach.
\subsubsection{Derivation of Likelihood Function}\label{appendixB1}
We first rewrite  \eqref{eqsignal10} as
$
{{{\bm{\mathord{\buildrel{\lower3pt\hbox{$\scriptscriptstyle\frown$}}
\over Y} }}}_b}= {\bm{A}}{\bm P}_b + {{\bm U}_{b}},
$,
where    ${{\bm{P}}_{b}} = \sum\nolimits_{k = 1}^K {{{\left( {{\bm{X}}_k^{:,{\Omega _{\rm{D}}} }} \right)}^H}{\bm{A}}_R^H{{\bm{v}}_b}{{{{\bm{s}}_k}}}}\in{\mathbb C}^{N_f\times T}$ and ${{\bm{A}}} = {\bm{A}}_T^{:,{\Omega _{\rm{D}}}}\in {\mathbb C}^{M\times N_f}$ for notation simplicity. Using the maximum likelihood approach, the parameters $\bm{A}$ and $\bm{P}_b$ are deterministic unknowns that we wish to estimate.% Keep in mind to estimate the span of $\mathbf{A}$, so $\bm{P}_b$ are actually nuisance parameters.

Next, we derive the likelihood function of $\bm Y$ in \eqref{eqsignal10v2} given the parameters ${{\bm{A}}}$ and $\bm{P}_b$.
Since $\bm Y$ is Gaussian distributed with fixed ${{\bm{A}}}$ and $\bm{P}_b$, the likelihood function is given by
\begin{align}
&{\cal P}({\bm{Y}}\left| {{\bm{A}},{{\bm{P}}_b}} \right.)= \frac{1}{{{\pi ^{MBT}}{\delta ^{2MBT}}}}\times \nonumber\\
 &
 {\exp \left( { - \frac{1}{{{\delta ^2}}}\sum\limits_{b = 1}^B {\sum\limits_{t = 1}^T {{{\left( {{\rm{ }}{\bm{\mathord{\buildrel{\lower3pt\hbox{$\scriptscriptstyle\frown$}}
\over Y} }}_b^{:,t} - {\bm{AP}}_b^{:,t}} \right)}^H}\left( {{\rm{ }}{\bm{\mathord{\buildrel{\lower3pt\hbox{$\scriptscriptstyle\frown$}}
\over Y} }}_b^{:,t} - {\bm{AP}}_b^{:,t}} \right)} } } \right)}.
\end{align}
Thus, the log-likelihood function is given by
\begin{align}
 & \log \left( {{\cal P}\left( {{\bm{Y}}\left| {{\bm{A}},{{\bm{P}}_b}} \right.} \right)} \right) =- MBT\log \left( {\pi {\delta ^2}} \right)  \nonumber\\
 & + \frac{1}{{{\delta ^2}}}\sum\limits_{b = 1}^B {\sum\limits_{t = 1}^T {{{\left( {{\rm{ }}{\bm{\mathord{\buildrel{\lower3pt\hbox{$\scriptscriptstyle\frown$}}
\over Y} }}_b^{:,t} - {\bm{AP}}_b^{:,t}} \right)}^H}\left( {{\rm{ }}{\bm{\mathord{\buildrel{\lower3pt\hbox{$\scriptscriptstyle\frown$}}
\over Y} }}_b^{:,t} - {\bm{AP}}_b^{:,t}} \right)} }.  \label{eqlast0}
\end{align}
For any given $\bm A$, the maximum likelihood estimate of each of the $\bm{P}_b^{:,t}$ is obtained by solving $
\frac{{\partial \log \left( {{\cal P}\left( {{\bm{Y}}\left| {{\bm{A}},{{\bm{P}}_b}} \right.} \right)} \right)}}{{\partial {\bm{P}}_b^{:,t}}} = 0,$
which gives $
{\bm{ P}}_b^{:,t} = {\left( {{{\bm{A}}^H}{\bm{A}}} \right)^{ - 1}}{{\bm{A}}^H}{\bm{\mathord{\buildrel{\lower3pt\hbox{$\scriptscriptstyle\frown$}}
\over Y} }}_b^{:,t}.$

Then, the log-likelihood as function of $\bm{A}$ is
\begin{align}
\log \left( {{\cal P}\left( {{\bm{Y}}\left| {\bm{A}} \right.} \right)} \right)=& - \frac{1}{{{\delta ^2}}}{\rm Tr}\left( {\left( {{{\bm{I}}_M} - {\bm{A}}{{\left( {{{\bm{A}}^H}{\bm{A}}} \right)}^{ - 1}}{{\bm{A}}^H}} \right){\bm{\hat C}}} \right)  \nonumber\\
 & - MBT\log \left( {\pi {\delta ^2}} \right), \label{eqlast2}
\end{align}
where  ${\bm {\hat C}}=\frac{1}{{BT}}{\bm{Y}}{{\bm{Y}}^H}$.

\subsubsection{Derivation of Subspace Estimation}\label{appendixB2}

Maximum likelihood estimate of $\bm{A}$ proceeds with maximizing \eqref{eqlast2} with resect to $\bm{A}$, i.e.,
\begin{align}
\mathop {\max }\limits_{\bm{A}} \;{\rm{Tr}}\left( {{\bm{A}}{{\left( {{{\bm{A}}^H}{\bm{A}}} \right)}^{ - 1}}{{\bm{A}}^H}{\bm{\hat C}}} \right)\label{eqlast3}.
\end{align}
Let ${\bm W}_\parallel{\bm\Delta} {\bm W}_\parallel^H$ be the eigenvalue decomposition of ${{\bm{A}}{{\left( {{{\bm{A}}^H}{\bm{A}}} \right)}^{ - 1}}{{\bm{A}}^H}}$, and ${{\bm\Delta} }={\rm diag}({\Delta _1},{\Delta _2},\cdots,{\Delta _{N_f}})$ be the eigenvalue matrix where the eigenvalues $\Delta_n$ are ordered in a decreasing order in magnitude.
Then, we know  ${\rm span}({\bm A})$ equals to the span of  ${\bm W}_\parallel\in{\mathbb C}^{M\times N_f}$, i.e.,
\begin{align}
{\rm span}({\bm A}) = {\rm span}({\bm W}_\parallel).\label{eqappendix5}
\end{align}
As mentioned, we only need to estimate the subspace ${\bm W}_\parallel$ without requiring exact AoD vectors $\bm A$. Hence, based on \eqref{eqappendix5}, we can reformulate \eqref{eqlast3} as
\begin{align}
\mathop {\max }\limits_{{\bm W}_\parallel} \;{\rm{Tr}}\left({\bm\Delta} {{\bm{W}}_{\parallel}^H{{\bm{S}}{\bm\Theta} {{\bm{S}}^{{H}}}}} {\bm W}_\parallel\right)+{\rm constant},\label{eqappendixy7}
\end{align}
where  ${{\bm{S}}{\bm\Theta} {{\bm{S}}^{{H}}}}$ is the eigenvalue decomposition of ${\bm {\hat C}}=\frac{1}{{BT}}{\bm{Y}}{{\bm{Y}}^H}$, and ${{\bm\Theta} }={\rm diag}([{\theta _1},{\theta _2},\cdots,{\theta _{M}}])\in{\mathbb C}^{M\times M}$ is the eigenvalue matrix where the eigenvalues $\theta_m$ are \mbox{ordered} in a decreasing order in magnitude.
Denote ${a_{n,m}} = {| {( {{\bm{W}}_\parallel ^{:,n}} )^H{{\bm{S}}^{:,m}}} |^2}$  and ${\tau _{n,m}} = \Delta_n{\theta _m}$ for $1\le n \le N_f$ and $1\le m \le M $.  Ignoring the constant term in \eqref{eqappendixy7}, we have
 \begin{align}
\mathop {\max }\limits_{ {a_{n,m}\ge0}}\quad &  \sum\nolimits_{n = 1}^{{N_f}}{\sum\nolimits_{m = 1}^M {{a_{n,m}}\tau_{n,m}} }\nonumber\\
{\rm s.t.}\quad
&\sum\nolimits_{m = 1}^M {{a_{n,m}} = 1,\quad1\le n \le N_f,}\label{eqeq64b}\\
&\sum\nolimits_{n = 1}^{{N_f}} {{a_{n,m}} \le 1,}\quad1\le m \le M, \label{eqeq64}
%&\;a_{n,m}\ge0, \quad1\le n \le N_f, 1\le m \le M,\label{eqeq64a}
 \end{align}
where constraints \eqref{eqeq64b} and \eqref{eqeq64} come from the following properties of unitary matrices:
  \begin{align}
  {{\bm{S}}^{{H}}}\left[ {\begin{array}{*{20}{c}}
{{{\bm{W}}_\parallel }},{{{\bm{W}}_ \bot }}
\end{array}} \right]\left[ {\begin{array}{*{20}{c}}
{{\bm{W}}_\parallel ^{{H}}}\\
{{\bm{W}}_ \bot ^{{H}}}
\end{array}} \right]{\bm{S}} = {{\bm{I}}_M}, \nonumber\\
\left[ {\begin{array}{*{20}{c}}
{{\bm{W}}_\parallel ^{{H}}}\\
{{\bm{W}}_ \bot ^{{H}}}
\end{array}} \right]{\bm{S}}{{\bm{S}}^{{H}}}\left[ {\begin{array}{*{20}{c}}
{{{\bm{W}}_\parallel }}, {{{\bm{W}}_ \bot }}
\end{array}} \right] = {{\bm{I}}_M},
 \end{align}
where ${\bm W}_\bot$ denotes the $(M-N_f)$ orthonormal bases \mbox{corresponding} to the null space of $\bm{A}$. We know that problem \eqref{eqeq64} is a linear programming problem. Due to that $\Delta_1 \ge \Delta_2 \ge \cdots \ge \Delta_{N_f}$ and $\theta_1 \ge \theta_2 \ge \cdots \ge \theta_M$, and that $\tau_{n,m} = \Delta_n \theta_m$, the optimal solution of the linear program is given by
 \begin{align}
{a_{n,m}} = \left\{ {\begin{array}{*{20}{l}}
{1,\qquad {\rm if}\; 1 \le n = m \le {N_f},}\\
{0, \qquad {\rm otherwise}.}
\end{array}} \right.
 \end{align}
Note that $a_{n,m}=1$ implies ${\bm{W}}_\parallel^{:,n}={\bm S }^{:,m}$, which indicates
${{\bm{W}}_\parallel } = {\bm S}_{\parallel} \buildrel \Delta \over =  \left[ {{{\bm{S}}^{:,1}},{{\bm{S}}^{:,2}},\cdots,{{\bm{S}}^{:,N_f}}} \right].$

\ifCLASSOPTIONcaptionsoff
  \newpage
\fi
  \bibliography{SPT}
\bibliographystyle{IEEEtran}%By using IEEEtrans, the number can be displayed.

\end{document}